\documentclass[preprint]{aastex631}

\usepackage{newtxtext,newtxmath,color,enumerate}
\usepackage{amsmath, graphics, bm, graphicx, longtable, array, comment}

\begin{document}
\title{Linear Thermal Instability of a Condensing Gas-Particle
  Mixture, with Possible Application to Chondrites and Planetesimals}

\author[0009-0000-3886-8014]{Kecheng Stephon Qian}\thanks{e-mail:kq38@berkeley.edu}
\affiliation{Astronomy Department, Theoretical Astrophysics Center, and Center for Integrative Planetary Science, \\University of California, Berkeley, Berkeley, CA 94720, USA \\}

\author[0000-0002-6246-2310]{Eugene Chiang}\thanks{e-mail:echiang@astro.berkeley.edu}
\affiliation{Astronomy Department, Theoretical Astrophysics Center, and Center for Integrative Planetary Science, \\University of California, Berkeley, Berkeley, CA 94720, USA \\}
\affiliation{Department of Earth and Planetary Science, University of California, Berkeley, CA 94720, USA}



\begin{abstract}
We study the stability of a hot saturated gas coexisting with condensed particles in an optically thin medium. Such a situation may obtain downstream of a shock, at condensation fronts, or in vaporizing impacts. We show that the gas-particle mixture is subject to a thermal instability whereby a region of lower temperature and higher condensate density cools faster to condense faster. If the region of runaway condensation has a sound-crossing time shorter than its cooling time, then it accretes more mass, in gas and particles, from its higher pressure surroundings. Numerical integration of the linearized perturbation equations demonstrates that this radiation-condensation instability can create particle clumps and voids out of a secularly cooling gas. Provided radiation can escape to cool particle overdensities, thermal instability can help assemble chondrite parent bodies out of the vaporized debris of asteroid collisions, and form planetesimals generally.
\end{abstract}


\section{Introduction}\label{sec:intro}
Condensation fronts, where gas condenses into liquid or solid particles, appear in many contexts. In planetary atmospheres, clouds condense out of water vapor (Earth), sulfuric acid (Venus), carbon dioxide (Mars), ammonia (Jupiter), methane (Uranus), and silicates and iron (brown dwarfs and hot Jupiters; for a introduction to the microphysics of clouds, see \citealt{pruppacher_klett_2010}). Water vapor and CO freeze out where molecular clouds and protoplanetary disks are sufficiently cold (``snowlines''; \citealt{qi_etal_2013}; \citealt{cieza_etal_2016}; \citealt{owen_2020}; \citealt{wang_etal_2025}). Dust of diverse mineralogies condenses from the outflows of evolved stars \citep{tielens_2022} and catastrophically evaporating rocky planets \citep{bromley_chiang_2023}.

Collisions between solid bodies are another source of vapor condensates. Meteor impacts have showered the lunar and terrestrial landscapes with silicate spherules condensed from impact vapor plumes \citep{johnson_melosh_2012,johnson_melosh_2012b,johnson_melosh_2014}. Of particular interest here are CB/CH chondritic meteorites which are nearly completely filled with mm-sized, once-liquid metal nodules and silicate chondrules. These melt droplets are thought to be condensed from vaporizing collisions of differentiated asteroids (\citealt{choksi_etal_2021}, and references therein; see also \citealt{stewart_etal_2025}). Impact plumes from colliding rocky bodies are special because they present a wholly condensable medium of hot rock and metal vapor, undiluted by inert, non-condensable gases like hydrogen. Such ``second-generation'' gas may also be found in extrasolar debris disks of colliding planetesimals, around young stars (e.g.~\citealt{marino_etal_2022}) and white dwarfs (e.g.~\citealt{swan_etal_2023}).

How do fresh melt droplets from an explosion agglomerate into CB/CH chondrite meteorites? The droplets must re-collect promptly and efficiently to explain their nearly 100\% volume-filling fractions. A proposed solution to the ``Humpty-Dumpty'' problem is a radiation-condensation instability \citep{chiang_2024}. The idea is that in a saturated vapor, regions overdense in particle condensates radiatively cool relative to their surroundings, losing pressure and collapsing into smaller volumes. The plume may fragment into cool, dense clumps of particles surrounded by hot, rarefied vapor, analogous to how thermal instability fragments the interstellar medium into multiple phases (\citealt{field_1965}; \citealt{jennings_li_2021}, and references therein). \citet{chiang_2024} investigated potential nonlinear outcomes in a saturated cloud of silicate vapor by modeling collapsing regions as cavitating bubbles.

We seek here to place the hypothesized radiation-condensation instability on firmer ground by seeing if and how it emerges from a linear stability analysis. We consider an initially uniform density medium composed of a wholly condensable, saturated gas and its entrained particle condensates, and ask whether small disturbances to this fluid grow or damp. Working in the optically thin limit where radiation from particles is allowed to escape to infinity, we will indeed find fast-acting instabilities that grow particle overdensities. Our study is the linear counterpart to the nonlinear explorations of \citet{chiang_2024}, and offers, like the streaming instability and other particle concentration mechanisms (see \citealt{chiang_youdin_2010} for a review), a means of creating overdensities that can lead to planetesimal formation.

Stability analyses usually presume a fixed background equilibrium state to perturb. In Section \ref{sec:scientific}, to define such an equilibrium, we introduce an arbitrary and fixed source of background heating to balance radiative cooling from background particles. This equilibrium is perturbed to derive an analytic dispersion relation for Fourier modes. We work out modal growth rates, and physically interpret mode behaviors. The goal of this analytic section is to develop physical intuition for the radiation-condensation instability, in the hope that some of the behaviors uncovered will be robust against our use of an artificial heating term in the energy equation.

Section \ref{sec:numexp} solves the linear perturbation equations numerically. After validating our integrator by reproducing our analytic eigenmodes, we conduct experiments that remove the restrictions of our analytic study --- in particular, we dispense with background heating and allow the medium to secularly cool by radiation. This time-dependent background is more physically realistic, as secular cooling better describes the evolution of a collisional plume, or the fluid downstream of a shock front. 
On top of this time-dependent background we introduce linear perturbations and study their growth. 

Section \ref{sec:sum} summarizes and places the radiation-condensation instability in the context of Field's (\citeyear{field_1965}) thermal instability.

\section{Linear stability analysis of a fixed background}
\label{sec:scientific}

We assess the linear stability of a gas-particle mixture, where the two species inter-convert through phase changes, are tightly coupled by mutual drag, and the particles radiate freely to space. No magnetic fields, conduction/diffusion, viscosity, background velocities including turbulence, are considered. We carry out the usual analytic procedure of Fourier analyzing perturbations to a background state. Fourier analysis and the derivation of a wave dispersion relation require that the perturbations vary smoothly in space and time, and that the background be in a time-independent equilibrium (so that the linear algebraic perturbation equations used to derive the dispersion relation have constant coefficients).  As mentioned in \S\ref{sec:intro}, to construct such an equilibrium, we will need to introduce a background heating term into the energy equation, to balance radiative losses from background particles. This artifice enables us to analytically survey and explore a wide range of physical behaviors, some of which will hopefully still manifest in more realistic set-ups without background heating. We will comment on which effects may be robust and which effects may not be (see \S\ref{subsec:stat_eigen} and \S\ref{subsec:acou_eigen}), and test our assertions against numerical experiments in \S\ref{sec:numexp}.

The equations governing our fluid mixture are introduced in their most basic form in \S\ref{subsec:mme}. The background equilibrium state is described in \S\ref{subsec:back}. Linear perturbation equations are derived in \S\ref{subsec:linper}, and solved for eigenfrequencies in \S\ref{subsec:ef} and eigenmodes in \S\ref{subsec:em}.

\subsection{Mass, momentum, and energy equations}\label{subsec:mme}
Consider a condensable gas mixed with its liquid or solid particle condensates, of total mass density $\rho_{\rm tot} = \rho_{\rm gas} + \rho_{\rm par}$, for gas density $\rho_{\rm gas}$ and particle density $\rho_{\rm par}$. We do not distinguish between solid and liquid phases for the particles. At the time of their formation, chondrules and metal nodules from CB chondrites were at least partially liquid while suspended in space, to attain their observed spherical shapes.

Gas and particles are assumed to be in thermal and chemical equilibrium: on the pressure $P$ vs.~temperature $T$ phase diagram, the mixture is assumed to reside on the co-existence curve, such that the gas pressure is always given by the saturation vapor pressure
\begin{align}
P = P_{\rm sat}(T)\,.
\end{align}
For molten ``bulk silicate earth'' having a composition similar to olivine-rich chondrites,
\begin{align}
\log_{10}\left(\frac{P_{\rm sat}}{\rm bars}\right) &= -30.6757 - \frac{8228.146\,\,{\rm K}}{T}+\, 9.3974\log_{10}\left(\frac{T}{{\rm K}}\right) 
    \label{eq:psat}
\end{align}
(\citealt{fegley_schaefer_2012}; for vapor pressures of other refractory materials, see \citealt{visscher_fegley_2013, perezbecker_chiang_2013}). The vapor behaves as an ideal gas,
\begin{align} \label{eq:eos}
P = \frac{\rho_{\rm gas} k_{\rm B} T}{\mu m_{\rm H}} \,,
\end{align}
so its density also depends on $T$ only,
\begin{align}
\rho_{\rm gas} = \rho_{\rm sat} (T) = \frac{\mu m_{\rm H}}{k_{\rm B}} \frac{P_{\rm sat}(T)}{T} 
\end{align}
for Boltzmann constant $k_{\rm B}$, mean molecular weight $\mu \simeq 30$, and atomic hydrogen mass $m_{\rm H}$. 

Gas and particles are assumed to be well-coupled dynamically (gas drag stopping times for chondrule-sized particles are suitably short in collisional vapor plumes; \citealt{choksi_etal_2021}), so that the two species move at a common velocity $\mathbf{v}$. The equations of mass and momentum evolution are given by
\begin{align}
\label{Eq:rho_tot}
\frac{D\rho_{\rm tot}}{Dt} 
&=-\rho_{\rm tot} \mathbf{\nabla \cdot v} \\
\label{Eq:rho_gas} 
\frac{D\rho_{\rm gas}}{Dt} 
&=  \frac{d\rho_{\rm sat}}{dT} \frac{DT}{Dt} \\
\label{eq:mom}
\frac{D\mathbf{v}}{Dt} 
&=-\frac{{1}}{\rho_{\rm tot}}\mathbf{\nabla} P
\end{align}
where $D/Dt \equiv \partial/\partial t + (\mathbf{v \cdot \nabla})$ is the Lagrangian derivative. 

Equation (\ref{Eq:rho_tot}) is the usual one for mass continuity; the total density of a fluid parcel (of fixed total mass) changes only by changing the parcel's volume, via the velocity divergence $\mathbf{\nabla \cdot v}$. The velocity divergence is absent from the gas continuity equation (\ref{Eq:rho_gas}) because we have assumed $\rho_{\rm gas} = \rho_{\rm sat}(T)$; the saturated gas density of a fluid parcel can only change from temperature changes, and not from volume changes per se. It follows that the particle density of a parcel can change from various effects:
\begin{align} 
\frac{D\rho_{\rm par}}{Dt} = \frac{D\rho_{\rm tot}}{Dt} - \frac{D\rho_{\rm gas}}{Dt} & = -\rho_{\rm tot}\mathbf{\nabla \cdot v} - \frac{d\rho_{\rm sat}}{dT} \frac{DT}{Dt} \nonumber \\
&= -\rho_{\rm par}\mathbf{\nabla \cdot v} - \rho_{\rm gas}\mathbf{\nabla \cdot v} - \frac{d\rho_{\rm sat}}{dT} \frac{DT}{Dt}\,.
\label{eq:rhopar}
\end{align}
The first term on the right-hand side of (\ref{eq:rhopar}) describes how the particle density changes from parcel volume changes at fixed particle mass (no phase changes). The second term describe how gas and particles can inter-convert from parcel volume changes at fixed temperature. The same phenomenon is evident in a piston enclosing vapor and liquid in equilibrium; at fixed temperature, lowering the piston to shrink the enclosed volume converts vapor to liquid while keeping the vapor density and pressure constant (see any thermodynamics textbook; e.g., Figure 10.1 of \citealt{kittel_kroemer_1980}). The third term accounts for phase changes from temperature changes.

The momentum eq.~(\ref{eq:mom}) describes how the fluid accelerates from gas pressure, with the  inertia given by $\rho_{\rm tot}$ for our assumed well-coupled particle-gas mixture. Though the fluid may be orbiting a star, rotational forces and orbital shear are negligible as long as we focus on processes than unfold over timescales much shorter than an orbital period (the timescale over which Coriolis and stellar tidal forces act). Petrologic experiments constrain chondrules to cool over timescales of hours to days \citep[e.g.][]{desch_connolly_2002,hewins_etal_2018}, much less than a heliocentric orbital period at the location of the asteroid belt.

The last equation needed to close the system is the energy equation: 
\begin{align}
\rho_{\rm tot} C \frac{DT}{Dt} = -P \mathbf{\nabla \cdot v} + L_{\rm vap} \left(-\rho_{\rm gas} \mathbf{\nabla \cdot v} - \frac{d\rho_{\rm sat}}{dT} \frac{DT}{Dt} \right) - 4\sigma T^4 \rho_{\rm par} \kappa_{\rm par} + \mathcal{H}
\label{eq:energy}
\end{align}
where $C \simeq 3k_{\rm B}/(\mu m_{\rm H}) \simeq 8 \times 10^6$ erg/g/K is the specific heat of the particle-gas mixture (neglecting the order-unity difference between particle and gas specific heats), $L_{\rm vap} \simeq 3 \times 10^{10}$ erg/g is the latent heat of vaporization ({\citealt{nagahara_etal_1994,kimura_etal_2002}), $\sigma$ is the Stefan-Boltzmann constant, and $\kappa_{\rm par}$ is a grey opacity (emissive cross-section per unit particle mass) which depends on the particle size distribution. We adopt, for a single particle size $s = 0.1$ cm and internal particle density $\rho_{\rm p} \simeq 3$ g/cm$^3$,  a fiducial $\kappa_{\rm par} = \pi s^2/(4\pi \rho_{\rm p}s^3/3) = 2.5$ cm$^2$/g. 

From left to right on the right-hand side of the energy eq.~(\ref{eq:energy}), the temperature of a parcel can change from:
\begin{enumerate}[(i)]
\item $PdV$ work 
\item latent heat release from condensation; the parentheses enclose only those terms in $D\rho_{\rm par}/Dt$ that involve phase changes (thus the first term on the right-hand side of eq.~\ref{eq:rhopar} does not qualify)
\item energy loss from radiation, modeled by assigning to particles a blackbody volume emissivity $j_\nu = B_{\nu} \rho_{\rm par} \kappa_{\rm par}$ for Planck source function $B_\nu$. Self-absorption is ignored --- the background is assumed optically thin, so that all radiation escapes to infinity. 
The factor of $4$ arises from integrating the Planck function first over frequency $\nu$ (yielding $\sigma T^4/\pi$) and then over all solid angle (yielding $4\pi$). The assumption of an optically thin medium may be critical for thermal instability; see \S\ref{sec:sum} for further discussion.
\item a constant heating term $\mathcal{H}$, introduced to balance radiation losses and formally define a background equilibrium. 
\end{enumerate}




\subsection{Background equilibrium state} \label{subsec:back}
A heating term $\mathcal{H} > 0$ is needed to define a background
equilibrium temperature $T_0 = [\mathcal{H}/(4\sigma \rho_{\rm
  par,0}\kappa_{\rm par})]^{1/4} >0$. In all our calculations below,
we take $T_0 = 2300$ K as a fiducial --- this choice lies between condensation temperatures of $\sim$2600 K for CB chondrite metals  (\citealt{campbell_etal_2002}, their Fig.~6) and a minimum inferred vapor plume temperature of 1928 K based on skeletal olivine chondrules in CB chondrites (\citealt{hewins_etal_2018}). In reality a cooling metal + silicate vapor will condense different species at different temperatures; accounting for the condensation sequence in detail is left for future work.

The background equilibrium state (subscript 0) is spatially uniform ($T_0 =$ constant, $P_0 = P_{\rm sat}(T_0)=$ constant, $\rho_{\rm gas,0}=\rho_{\rm sat}(T_0)=$  constant, $\rho_{\rm par,0}=$ constant) and motionless ($\mathbf{v}_0 = \mathbf{0}$). Over the course of this paper, we will experiment with different values for $\rho_{\rm par,0}$, ranging from zero to $0.1 \rho_{\rm gas,0}$.

\subsection{Linear perturbation equations}\label{subsec:linper}
We now introduce perturbations in the form of one-dimensional plane-parallel waves. For example, for pressure, $P = P_0 + \delta P \exp i(kx - \omega t)$, where $\delta P$ is the complex wave amplitude (of magnitude $|\delta P| \ll P_0$), $k$ is the real wavenumber (of magnitude $2\pi$ divided by the wavelength), $x$ is one-dimensional position, and $\omega$ is the complex wave frequency. The complex exponential form of the perturbation is introduced for mathematical convenience; the physical content is contained in the real part of $P$. If the imaginary part of $\omega$ is positive (${\rm Im}(\omega)>0$), then the wave amplifies exponentially in time, i.e.~the fluid is unstable.

We substitute $P = P_0 + \delta P \exp i(kx - \omega t)$, $T = T_0 + \delta T \exp i(kx - \omega t)$, $\mathbf{v} = v_0 \mathbf{\hat{x}} + \delta v \mathbf{\hat{x}} \exp i(kx-\omega t)$, etc., into the evolutionary equations (\ref{eq:eos}, \ref{Eq:rho_tot}, \ref{Eq:rho_gas}, \ref{eq:mom}, and \ref{eq:energy}; the condition of saturation equilibrium has been folded into equations \ref{Eq:rho_gas} and \ref{eq:energy}). After subtracting off the zeroth-order background terms (including the assumed constant $\mathcal{H}$),
and keeping only terms linear in perturbed quantities, we arrive at a homogeneous set of algebraic relations (subscript 0 dropped for convenience):
\begin{align}
\label{eq:drhotot}
-i\omega \,\delta \rho_{\rm tot}  =& -ik\rho_{\rm tot} \,\delta v \\
\label{eq:drhogas} 
-i\omega \,\delta \rho_{\rm gas} =& - i\omega \frac{d\rho_{\rm sat}}{dT} \,\delta T \\
\label{eq:dv}
-i\omega \,\delta v =&-\frac{{ik}}{\rho_{\rm tot}} \,\delta P\\
    \label{eq:de}
    -i \omega \rho_{\rm tot}C \, \delta T=& -ik P \,\delta v  -i k \rho_{\rm gas} L_{\rm vap} \, \delta v + i \omega L_{\rm vap} \frac{d\rho_{\rm sat}}{dT} \,\delta T - 
    4 \sigma T^4 \kappa_{\rm par} \,\delta \rho_{\rm par} \nonumber \\
     &- 16 \sigma T^3 \rho_{\rm par}
     \kappa_{\rm par} \, \delta T \\
    \frac{\delta P}{P} =& \frac{\delta \rho_{\rm gas}}{\rho_{\rm gas}} + \frac{\delta T}{T} \,. \label{eq:deos}
\end{align}
Note in (\ref{eq:de}) we have not assumed $|\delta \rho_{\rm par}| / \rho_{\rm par} \ll 1$; only the background terms have been subtracted, and $|\delta T|/T \ll 1$ assumed to keep terms linear in $\delta T$. 
We will be interested in the case where the background is nearly all gas ($\rho_{\rm tot} \simeq \rho_{\rm gas} \gg \rho_{\rm par}$), in which case $\rho_{\rm par}$ may be so small that $|\delta T|/T \ll |\delta \rho_{\rm par}|/\rho_{\rm par}$. Accordingly, for simplicity, we drop  
the last term of (\ref{eq:de}) relative to the second-to-last term:
\begin{align}
-i \omega \left(  \rho_{\rm tot}C + L_{\rm vap} \frac{d\rho_{\rm sat}}{dT} \right) \, \delta T&= -ik \left(P+\rho_{\rm gas} L_{\rm vap}\right) \delta v -
    4 \sigma T^4 \kappa_{\rm par} \left(\delta \rho_{\rm tot} -\delta \rho_{\rm gas}\right) \label{eq:dea} \tag{13a}
    \end{align}
where $(\delta \rho_{\rm tot} - \delta \rho_{\rm gas}) = \delta \rho_{\rm par}$. The dropped term $- 16 \sigma T^3 \rho_{\rm par} \kappa_{\rm par}  \delta T$ will be restored in the more accurate numerical experiments of \S\ref{sec:numexp}.

To re-cap the small parameters: $|\delta T|/T$, $|\delta P|/P$, $|\delta \rho_{\rm gas}|/\rho_{\rm gas} \ll 1$. We have not assumed $|\delta \rho_{\rm par}| \ll \rho_{\rm par}$ 
or $|\delta v| \ll v$ (the background $v = 0$). There are further restrictions from mass conservation. Since $\rho_{\rm par} + \delta \rho_{\rm par} \geq 0$ and $\rho_{\rm gas} + \delta \rho_{\rm gas} \geq 0$ (no negative masses), the perturbation densities have ``floors'':
\begin{subequations} \label{eq:bounds0}
\begin{align} \label{eq:bounds}
\delta \rho_{\rm par} & \geq -\rho_{\rm par} \\
\label{eq:boundsa} \delta \rho_{\rm gas} & \geq -\rho_{\rm gas} \,.
\end{align}
\end{subequations}
Furthermore, the contribution to $\delta \rho_{\rm par}$ from phase changes alone (i.e.~not counting the contribution from particle transport) 
must be $\leq \rho_{\rm gas}$, as one cannot condense more than what is available in background gas. Likewise $\delta \rho_{\rm gas}$ from phase changes alone must be $\leq \rho_{\rm par}$. Thus there are also ``ceilings'':
\begin{subequations} \label{eq:bounds2}
\begin{align} \label{eq:bounds1}
\delta \rho_{\rm par,~phase~change~only} & 
\leq \rho_{\rm gas} \\ \label{eq:bounds1a}
\delta \rho_{\rm gas,~phase~change~only} & 
\leq \rho_{\rm par}\,.
\end{align}
\end{subequations}
Analytically (this section \ref{sec:scientific}), perturbations $\delta \rho_{\rm par}$ and $\delta \rho_{\rm gas}$ may always be scaled small enough to stay within the bounds (\ref{eq:bounds0})--(\ref{eq:bounds2}), as required by Fourier analysis (where all derivatives are continuous). In our numerical experiments (\S\ref{sec:numexp}), we will occasionally and by design hit up against the bounds, and enforce them manually.

In matrix form, our simplified linear perturbation equations are:
\begin{align}
\begin{bmatrix}
\omega  & 0 & - k \rho_{\rm tot} & 0 & 0 \\
0  & \omega & 0 & - \omega (d\rho_{\rm sat}/dT)  & 0  \\
0  & 0 & \omega & 0 & - ( k/\rho_{\rm tot} ) \\
4\sigma T^4 \kappa_{\rm par} & -  4 \sigma T^4 \kappa_{\rm par}  & ik (P + \rho_{\rm gas} L_{\rm vap}) & -i\omega \left[ \rho_{\rm tot} C + L_{\rm vap} (d\rho_{\rm sat}/dT) \right] 
& 0 \\
0 &  - 1/\rho_{\rm gas}  & 0  & -1/T  & 1/P 
\end{bmatrix}
\begin{bmatrix}
\delta \rho_{\rm tot} \\ \delta \rho_{\rm gas} \\ \delta v \\ \delta T \\ \delta P
\end{bmatrix}
= \mathbf{0}. \label{eq:matrix}
\end{align}


\vspace{0.1in}

\subsection{Dispersion relation and eigenfrequencies}\label{subsec:ef}

\vspace{0.1in}

Setting the determinant of the 5$\times$5 matrix in (\ref{eq:matrix}) equal to 0 gives the dispersion relation:
\begin{align}\label{eq:cubic}
(1 + \ell a) \, \omega^3 - ia \omega_{\rm T}\, \omega^2 - (1 + a) (b + \ell) c^2 k^2 \, \omega + i (1+a) \omega_{\rm T} c^2 k^2  = 0 
\end{align}
where we have defined a frequency
\begin{align}
\omega_T \equiv \frac{4\sigma T^4 \kappa_{\rm par}}{CT} \simeq 0.8 \left( \frac{T}{2300 \, {\rm K}} \right)^3 \left( \frac{\kappa_{\rm par}}{2.5 \, {\rm cm}^2/{\rm g}} \right) {\rm s}^{-1}\,,
\end{align}
an isothermal sound speed 
\begin{align}
c \equiv \sqrt{P/\rho} \simeq 0.8 \, \left( \frac{T}{2300 \, {\rm K}} \right)^{1/2} {\rm km}/{\rm s} \,,
\end{align}
and dimensionless constants
\begin{align}
a \equiv \frac{d\ln \rho_{\rm sat}}{d\ln T} = \frac{8228.146 \,{\rm K}}{T\log_{10}e } -1 + 9.3974 &\simeq 16.6 \label{eq:a} \\
b \equiv \frac{P}{\rho C T} &\simeq 0.3 \\
\ell \equiv \frac{L_{\rm vap}}{C T} &\simeq 1.6 \left( \frac{2300 \, {\rm K}}{T} \right)
\end{align}
all while assuming the background particle density is small compared to the background gas density
\begin{align} \label{eq:parnone}
\rho_{\rm par} \ll \rho_{\rm gas} = \rho_{\rm sat} = \rho_{\rm tot} \equiv \rho \,.
\end{align}

Figure~\ref{fig:eigenfrequency} shows the eigenfrequency solutions of the cubic (\ref{eq:cubic}) solved numerically. All modes are unstable, growing exponentially in time. In the following subsections we analytically sketch eigenfrequencies and eigenmodes in various limits to develop physical intuition.

\begin{figure}
    \centering
    \includegraphics[width=0.70\linewidth]{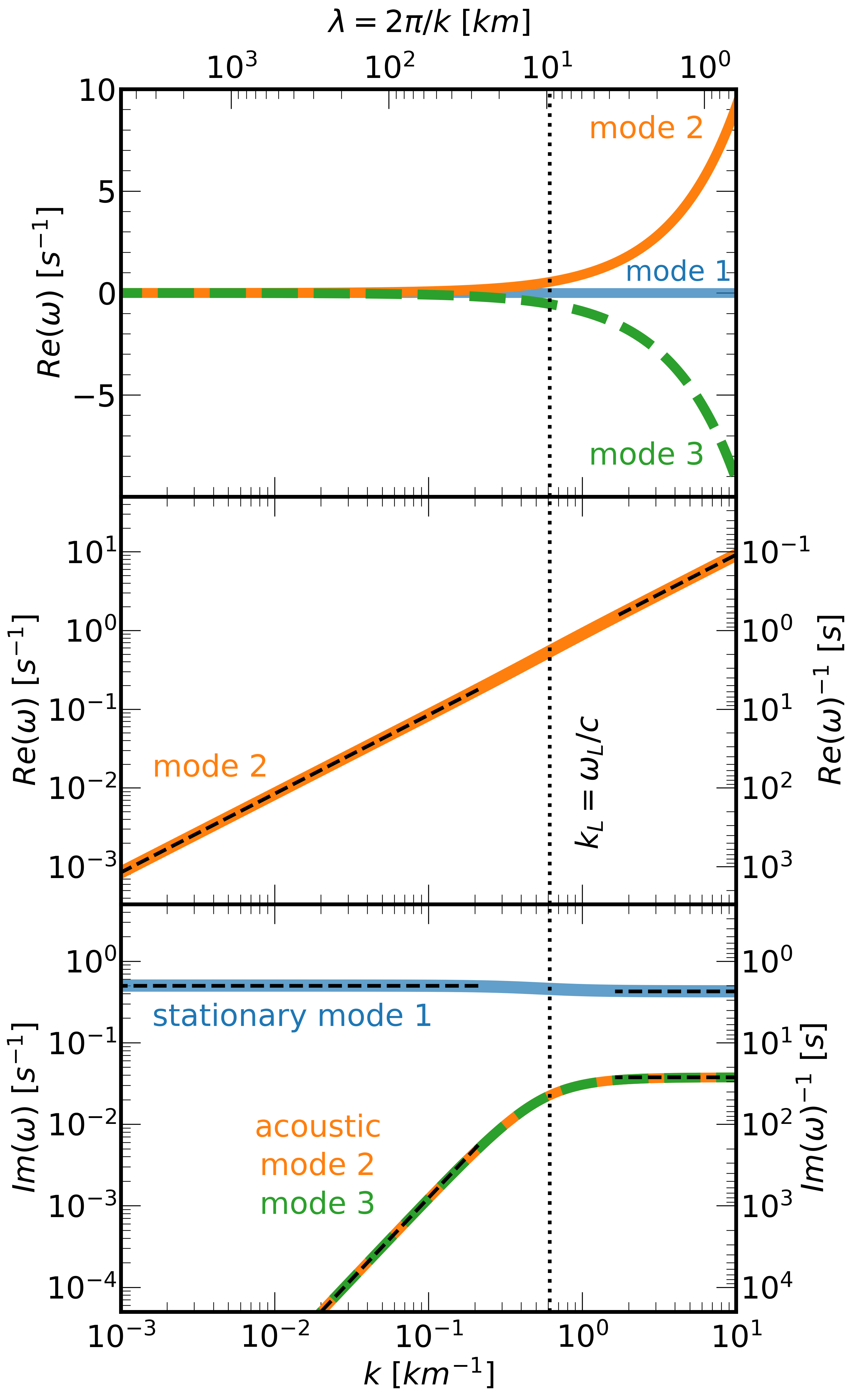}
    \caption{\small Eigenmode frequencies (colored curves), obtained by numerical solution of the simplified dispersion relation (\ref{eq:cubic}), evaluated for fiducial parameters $T = 2300$ K and $\kappa_{\rm par} = 2.5$ cm$^2$/g. Mode 1 is ``stationary'' [Re ($\omega_1) = 0$], while modes 2 and 3 are sound waves traveling in opposite directions at approximately speed $c \equiv \sqrt{P/\rho} \simeq 0.8$ km/s. All three modes grow with time [Im ($\omega) > 0$]. The dotted vertical line marks $k_{\rm L}= \omega_L/c$ dividing low-$k$ and high-$k$ regimes. Dashed black lines show analytic asymptotic results in those two regimes (eqs.~\ref{eq:stat1}, \ref{eq:mode23_highk}, and \ref{eq:mode23_lowk}). Agreement with the numerical results is excellent. 
    }
    \label{fig:eigenfrequency}
\end{figure}

\subsubsection{Stationary mode growth rates}
The alternating real and imaginary coefficients of the cubic (\ref{eq:cubic}) imply that a root with ${\rm Re}(\omega) = 0$ and ${\rm Im}(\omega) > 0$ --- a ``stationary'' (zero phase speed) growing mode --- is possible. 
We can estimate this purely imaginary root 
in high-$k$ and low-$k$ limits.  Dividing (\ref{eq:cubic}) by $(1+\ell a)$, and grouping terms to highlight stationary-mode behavior, we have 
\begin{align}
\omega^2 (\omega - i \omega_L) - c^2k^2 \left(\Gamma \omega - i\frac{1+a}{a}\omega_L\right) = 0 \label{eq:full-cub1}
\end{align}
where
\begin{align}
\omega_L \equiv \frac{a\omega_T}{1+\ell a} \simeq  \frac{4\sigma T^4 \kappa_{\rm par}}{L_{\rm vap}} \simeq 0.5 \left( \frac{T}{2300 \, {\rm K}} \right)^4 \left( \frac{\kappa_{\rm par}}{2.5 \, {\rm cm}^2/{\rm g}} \right)\, {\rm s}^{-1}
\end{align}
and
\begin{align}
\Gamma \equiv \frac{(1 + a)(b+ \ell)}{1+\ell a} \simeq 1.2. \label{eq:gamma}
\end{align}
For $k\rightarrow 0$, we can ignore the $c^2k^2$ term of (\ref{eq:full-cub1}), finding $\omega = +i\omega_L$. For $k\rightarrow \infty$, we keep only the $c^2k^2$ term, and find nearly the same result, $\omega = +i\omega_L(1+a)/(\Gamma a)$. 
We see that $ck_{\rm L} \simeq \omega_L$ divides the low-$k$ and high-$k$ limits. To summarize,
\begin{equation} \label{eq:stat1}
\omega_1 \approx \begin{cases} +i\omega_L & k \ll \omega_L/c \\
+i\omega_L (1+a)/(\Gamma a) & k \gg \omega_L/c \,.
\end{cases}
\end{equation}
Mode 1 is stationary and growing for all $k$. Equation (\ref{eq:stat1}) is verified by the full numerical solution of (\ref{eq:full-cub1}), as shown in Figure 1. We recognize mode 1 as the radiation-condensation instability hypothesized by \citet{chiang_2024}. It amplifies over the rate at which latent heat radiates away: $\omega_L \simeq \omega_T/\ell = 4\sigma T^4 \kappa_{\rm par}/L_{\rm vap}$.

\subsubsection{Acoustic mode frequencies and growth rates} \label{subsec:acoustic}
To find the other two roots of the cubic (\ref{eq:cubic}), we re-group terms again, this time highlighting (non-stationary) sound-wave behavior:
\begin{align}
\omega ( \omega^2 - \Gamma c^2 k^2) -i \omega_L \left(\omega^2 - \frac{1+a}{a} c^2 k^2\right) = 0.\label{eq:full-cub2}
\end{align}
We write $\omega = \omega_{\rm Re} + i \omega_{\rm Im}$, where the real part $\omega_{\rm Re} = \mathcal{O}(\pm ck)$ and the imaginary part $|\omega_{\rm Im}| \ll ck$. 
In the high-$k$ limit $\omega_L \ll ck$,  
 (\ref{eq:full-cub2}) is dominated by the first term and gives $\omega_{\rm Re} \simeq \pm \sqrt{\Gamma}ck$. Now insert $\omega = \pm \sqrt{\Gamma} ck + i\omega_{\rm Im}$ back into (\ref{eq:full-cub2}), dropping $\omega_{\rm Im}^2$ and $\omega_{\rm Im}\omega_L$ terms to solve for $\omega_{\rm Im}$ to leading order.
We find 
\begin{equation}
\omega_{2,3} \simeq \pm \sqrt{\Gamma} ck + i\omega_L(\Gamma-1-1/a)/(2\Gamma) \,\,\,\,\,\,\,\, k \gg \omega_L/c \,.
\label{eq:mode23_highk}
\end{equation}
In this high-$k$ limit, modes 2 and 3 are adiabatic sound waves (with adiabatic index $\Gamma$) that grow at rate  $\sim$$0.06\,\omega_L$. The high-$k$ growth rate is independent of $k$, and slower than for the stationary mode.

We now work in the $ck \ll \omega_L$ limit. The cubic (\ref{eq:full-cub2}) is then dominated by the $i\omega_L$ term, which yields $\omega \simeq \pm [(1+a)/a]^{1/2}ck$. We insert $\omega \simeq \pm (1+1/a)^{1/2}ck + i \omega_{\rm Im}$ 
into (\ref{eq:full-cub2}), dropping terms of order $\omega_{\rm Im}^2$ and $c^2k^2 \ll ck \omega_L$ to solve for $\omega_{\rm Im}$ to leading order. We find
\begin{equation}
\omega_{2,3} \simeq \pm (1+1/a)^{1/2}ck + \frac{i(\Gamma - 1-1/a)}{2} \frac{c^2k^2}{\omega_L} \,\,\,\,\,\,\,\, k \ll \omega_L/c \,.
\label{eq:mode23_lowk}
\end{equation}
In this low-$k$ limit, modes 2 and 3 are nearly isothermal sound waves that grow at rates that decrease with decreasing $k$. The low-$k$ and high-$k$ frequency behaviors for acoustic modes 2 and 3 are confirmed in Figure \ref{fig:eigenfrequency}.

\subsection{Eigenvectors and work integrals}\label{subsec:em}
To better understand the physical behaviors of the modes, we solve for the eigenvectors. We insert $\omega = \omega_{\rm Re}+i\omega_{\rm Im}$ 
into the matrix equation (\ref{eq:matrix}), and non-dimensionalize the state vector:
\begin{align}
\begin{bmatrix}
\rho (\omega_{\rm Re} + i\omega_{\rm Im})  & 0 & - \rho c k & 0 & 0 \\
0  & \rho (\omega_{\rm Re} + i\omega_{\rm Im}) & 0 & - a \rho (\omega_{\rm Re} + i\omega_{\rm Im})  & 0  \\
0  & 0 & c (\omega_{\rm Re} + i\omega_{\rm Im}) & 0 & -c^2 k \\
\ldots & \ldots & \ldots & \ldots & \ldots \\
0 &  - 1  & 0  & -1  & 1 
\end{bmatrix}
\begin{bmatrix}
\delta \rho_{\rm tot}/\rho \\ \delta \rho_{\rm gas}/\rho \\ \delta v /c \\ \delta T / T \\ \delta P / P
\end{bmatrix}
= \mathbf{0} \label{eq:matrix1}
\end{align}
where we replaced $\rho_{\rm gas} \simeq \rho_{\rm tot}$ (background is nearly particle-free) with $\rho$, and omitted specifying the 4th row of the square matrix because the remaining four rows suffice to solve for the eigenvector. 

\begin{figure}
    \centering
    \includegraphics[width=0.52\linewidth]{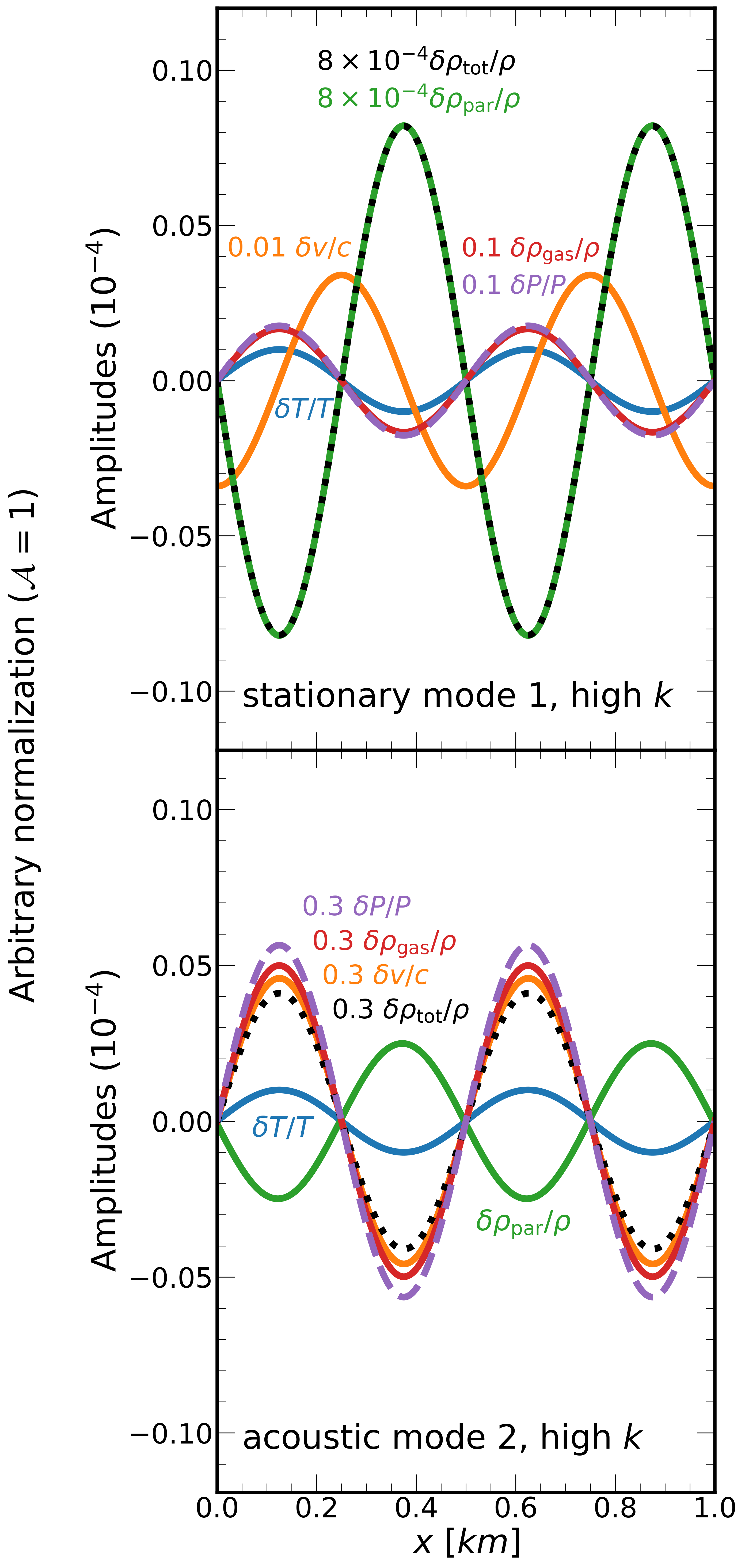}
    \hfill
    \includegraphics[width=0.469\linewidth]{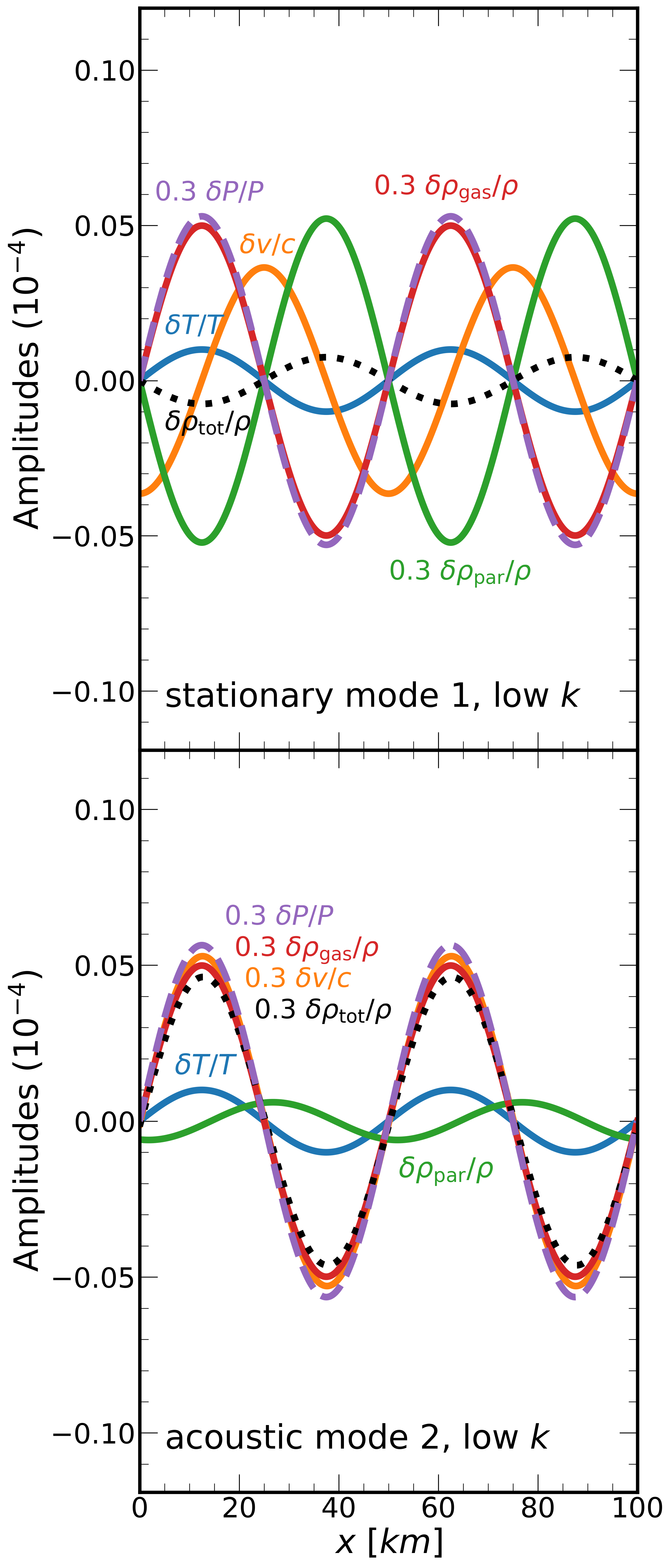}
  \caption{\small Eigenvector components of stationary mode 1 (having zero phase velocity; top panels), and of mode 2 which is a sound wave traveling in the positive $x$ direction (bottom panels). Modes are sampled at high $k$ (left) and low $k$ (right; note the change in $x$-scale) relative to $k_{\rm L} = \omega_L/c \simeq 0.6$ km$^{-1}$ (wavelength $2\pi/k_{\rm L} \simeq 10$ km). In every panel, the perturbation temperature $\delta T/T$ (blue curve) is assigned the same arbitrary amplitude and phase; amplitudes and phases of other perturbed quantities follow from the eigenmode, obtained by numerical solution of (\ref{eq:matrix1}). Note the annotated numerical coefficients, 
  introduced so that all curves fit in a given panel. The stationary mode at high $k$ (top left) has especially large velocities that concentrate particles especially strongly (see eq.~\ref{eq:eigen1}; we reiterate that this figure plots only eigenvector components scaled to the same $\delta T/T$, and shows no information about growth rates per se).}
    \label{fig:eigenmodes}
\end{figure}

\subsubsection{Stationary eigenmode behavior} \label{subsec:stat_eigen}
Stationary mode 1 has $\omega_{\rm Re} = 0$,   $\omega_{\rm Im} = \omega_L$ at low $k$, and $\omega_{\rm Im} = \omega_L(1+a)/(\Gamma a)$ at high $k$.
Without loss of generality we scale all eigenvector components to $\delta T/T$, and solve the equations in the 2nd, 5th, 3rd, and 1st rows of (\ref{eq:matrix1}), in that order, to find the stationary mode eigenvector
\begin{align}
\mathbf{E}_1 = \begin{bmatrix} \delta \rho_{\rm tot}/\rho \\ \delta \rho_{\rm gas}/\rho \\ \delta v /c \\ \delta T / T \\ \delta P / P
\end{bmatrix} = 
\begin{bmatrix}
-(ck/\omega_{\rm Im})^2 (1+a) \\ a \\ -i (ck/\omega_{\rm Im}) (1+a) \\ 1 \\ 1+a
\end{bmatrix}
\delta T/T \,. \label{eq:eigen1}
\end{align}
Pressure, gas density, and temperature fluctuations are all in phase with each other, with $\delta P$ and $\delta \rho_{\rm gas}$ of higher fractional amplitude than $\delta T$ (by about a factor of $a \simeq 17$) because of the exponential dependence of $P_{\rm sat}$ and $\rho_{\rm sat}$ on $T$. Velocity fluctuations $\delta v$ run ahead of pressure fluctuations $\delta P$ by a phase difference of $90^\circ$. At low $k$, $|\delta v/c| < |\delta P/P|$,
and at high $k$, $|\delta v/c| > |\delta P/P|$.
The particle density variation
\begin{align}
\frac{\delta \rho_{\rm par}}{\rho} = \frac{\delta \rho_{\rm tot} - \delta \rho_{\rm gas}}{\rho} = \left[-\left( \frac{ck}{\omega_{\rm Im}} \right)^2 \frac{1+a}{a} - 1\right] \frac{\delta \rho_{\rm gas}}{\rho} \label{eq:drhopar1}
\end{align}
is $180^\circ$ out of phase with the gas density variation, with $|\delta \rho_{\rm par}| = |\delta \rho_{\rm gas}|$ at low $k$, and $|\delta \rho_{\rm par}| > |\delta \rho_{\rm gas}|$ at high $k$.

Scaled pictures of the stationary eigenmode are shown in Figures~\ref{fig:eigenmodes}, \ref{fig:highk_mode1}, and \ref{fig:lowk_mode1}. Throughout this paper, we initialize the perturbation temperature to 
\begin{equation} \label{eq:norm}
\left. \frac{\delta T}{T} \right|_{t = 0} = \mathcal{A} \cdot 10^{-6} \sin kx
\end{equation}
where $\mathcal{A}$ is an arbitrary normalization constant, introduced for bookkeeping. Because our entire study is in the linear regime (including the numerical experiments of \S\ref{sec:numexp}), all perturbation quantities scale with $\mathcal{A}$, at least initially. For ease of comparison between different calculations, we set $\mathcal{A} = 1$ unless indicated otherwise.

In the low-$k$ limit (Figs.~\ref{fig:eigenmodes} and \ref{fig:lowk_mode1}), fluid velocities $\delta v$ and variations in total density $\delta \rho_{\rm tot}$ are negligible.  Cold and hot regions cool down and heat up independently, on timescale $\omega_{L}^{-1}$, before they can communicate by pressure disturbances which travel at speed $c$. Cold regions get colder because they condense more particles, which radiate more and cool the fluid more, in a positive feedback loop. Likewise hot regions get hotter because they have increasingly fewer particles. Changes in the local particle-to-gas ratio occur simply from local condensation and evaporation ($\delta \rho_{\rm par} \simeq - \delta \rho_{\rm gas}$).

By contrast, in the $ck \gg \omega_L$ limit (Figs.~\ref{fig:eigenmodes} and \ref{fig:highk_mode1}), hot and cold regions can communicate, and mass is transported between them. Most of the changes in total density are from the particle density changing ($|\delta \rho_{\rm par}| \gg |\delta \rho_{\rm gas}|$; eq.~\ref{eq:drhopar1}), as particles are transported out of high-pressure hot regions into low-pressure cold regions. Gas transports these particles, but gas densities do not rise in tandem with particle densities, because the gas density is throttled by saturation equilibrium; whatever gas moves with the particles into cold regions condenses into particles. In this short-wavelength limit, we have a material instability that collects increasing numbers of particles into colder, overdense clumps by excavating mass out of hotter voids.

How much of the stationary mode's behavior depends on our use of an artificial heating term $\mathcal{H}$? The heating of hot, particle-poor ($\delta \rho_{\rm par} < 0$) regions to temperatures above the background does depend on $\mathcal{H} > 0$ --- on the right-hand side of the energy equation (\ref{eq:energy}), heat exchange between the fluid and infinity arises from the net difference
\begin{align}
\mathcal{Q} \equiv \mathcal{H} - 4\sigma T^4 \kappa_{\rm par} \rho_{\rm par}
\end{align}
which is positive for $\delta \rho_{\rm par} < 0$. Conversely, in cold, particle-rich regions ($\delta \rho_{\rm par} > 0$), the difference $\mathcal{Q} < 0$. But $\mathcal{Q}$ can be still be negative in such regions if $\mathcal{H} = 0$; cold, particle-rich regions would still get colder relative to their surroundings by having more particles to emit more radiation. We are therefore led to believe that a growing mode similar to (but not identical to) the one we have found should exist when $\mathcal{H}=0$, driven by runaway cooling and condensation. We will test this assertion in \S\ref{sec:numexp}.

\begin{figure} 
    \centering
    \includegraphics[width=0.95\linewidth]{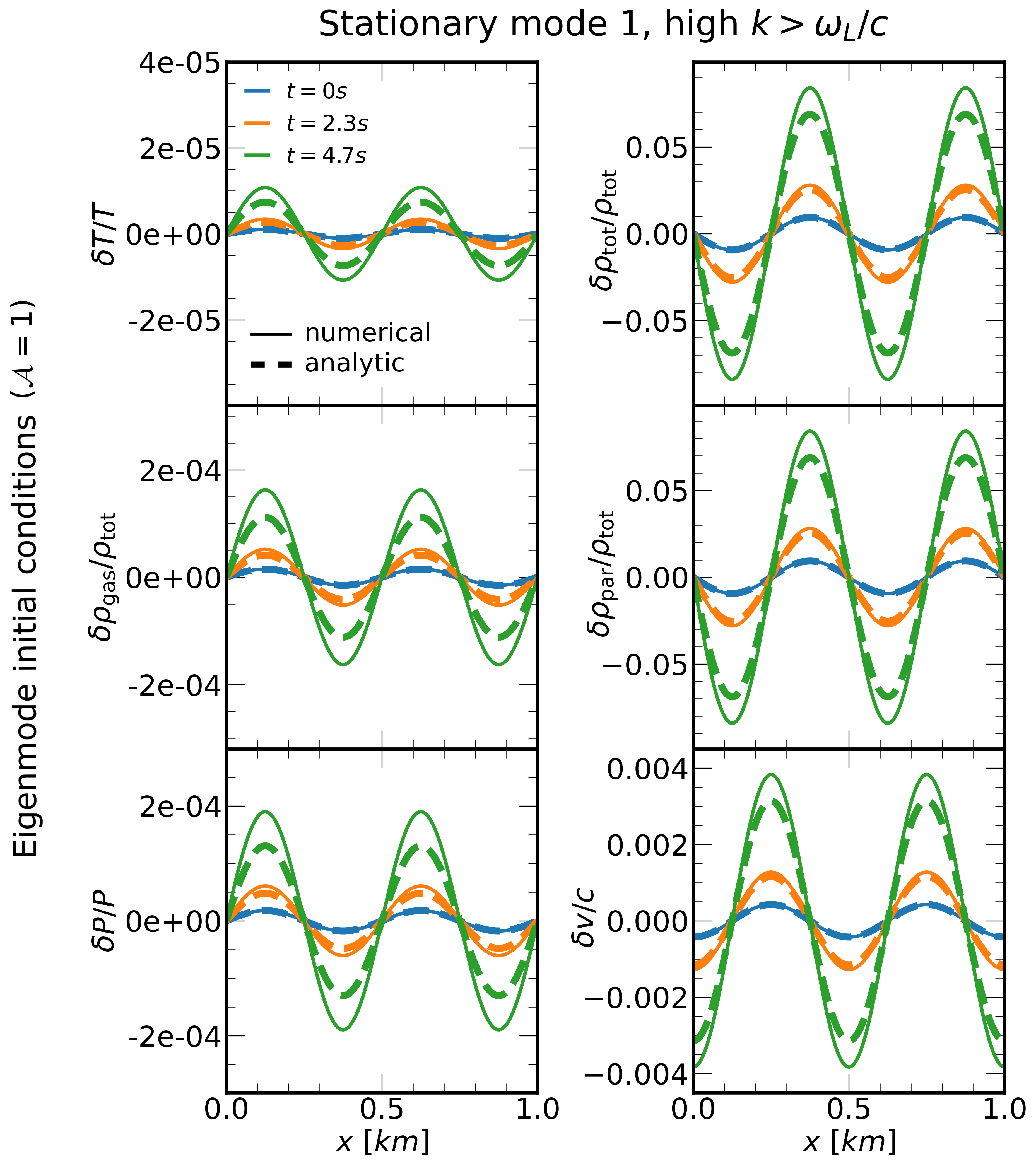}
    \caption{\small Time evolution of stationary mode 1 at high $k > \omega_L/c$, solved numerically using our staggered leapfrog integrator (thin solid curves, \S\ref{subsec:ot_eigen}), and overlaid with analytic solutions (thick dashed curves, \S\ref{subsec:stat_eigen}) obtained by multiplying initial values by $e^{-i \omega_1 t}$. Because the calculation is linear, all variables scale with a universal arbitrary constant $\mathcal{A}$ as defined in (\ref{eq:norm}) and chosen here to be 1. The relative magnitudes of variables (e.g.~$\delta \rho_{\rm par}/\rho_{\rm tot}$ vs.~$\delta \rho_{\rm gas}/\rho_{\rm tot}$) are fully determined. Note how $|\delta \rho_{\rm par}| \gg |\delta \rho_{\rm gas}|$; the mode has short enough wavelength that mass can move between hot and cold regions within a condensation time $\omega_L^{-1}$, increasing particle-to-gas ratios above what a purely condensing, static medium would yield (see Fig.~\ref{fig:lowk_mode1} for an approximation of the latter). The numerical solution adopts a background particle-to-gas ratio $\rho_{\rm par}/\rho_{\rm gas} = 0.1$, and deviates from the analytic which is computed in the limit $\rho_{\rm par}/\rho_{\rm gas} \ll 1$. For better agreement, we could reduce the background $\rho_{\rm par}$ in the numerical solution, except that $\delta \rho_{\rm par}$ grows so negative for this mode that it threatens to violate mass conservation, which requires $\rho_{\rm par} + \delta \rho_{\rm par} \geq 0$.} 
    \label{fig:highk_mode1}
\end{figure}

\begin{figure} 
    \centering
    \includegraphics[width=0.95\linewidth]{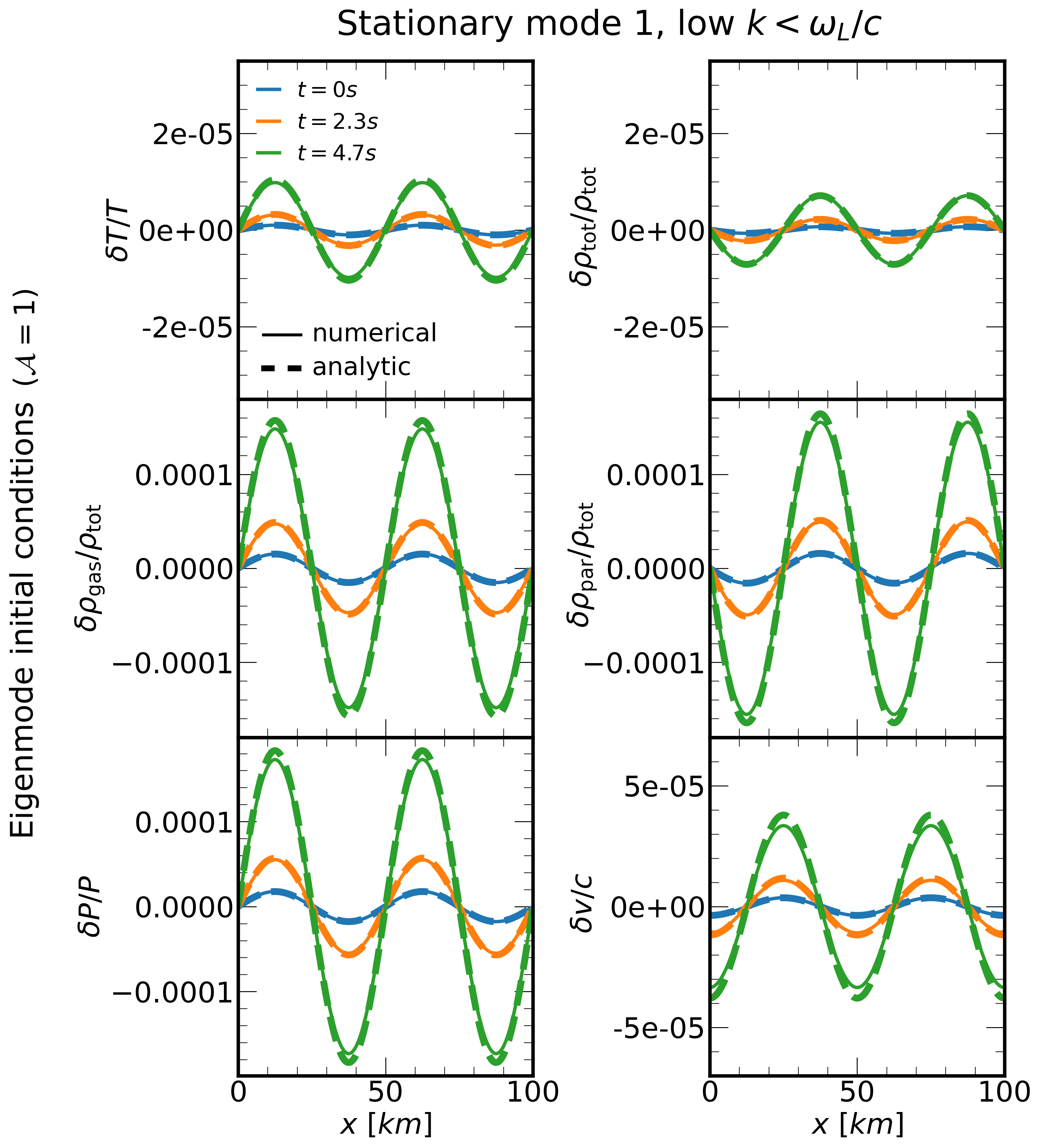}
    \caption{\small Same as Fig.~\ref{fig:highk_mode1}, except for the stationary mode at low $k$. 
    The mode has too long a wavelength for pressure gradients to transport much mass over the cooling time, and therefore gas mostly condenses without moving, with $\delta \rho_{\rm par} \simeq - \delta \rho_{\rm gas}$. As in Fig.~\ref{fig:highk_mode1}, the numerical solution adopts a background particle-to-gas ratio $\rho_{\rm par}/\rho_{\rm gas} = 0.1$, whereas the analytic curves are computed in the limit $\rho_{\rm par}/\rho_{\rm gas} \ll 1$.
    }
    \label{fig:lowk_mode1}
\end{figure}

\begin{figure}
    \centering
    \includegraphics[width=0.95\linewidth]{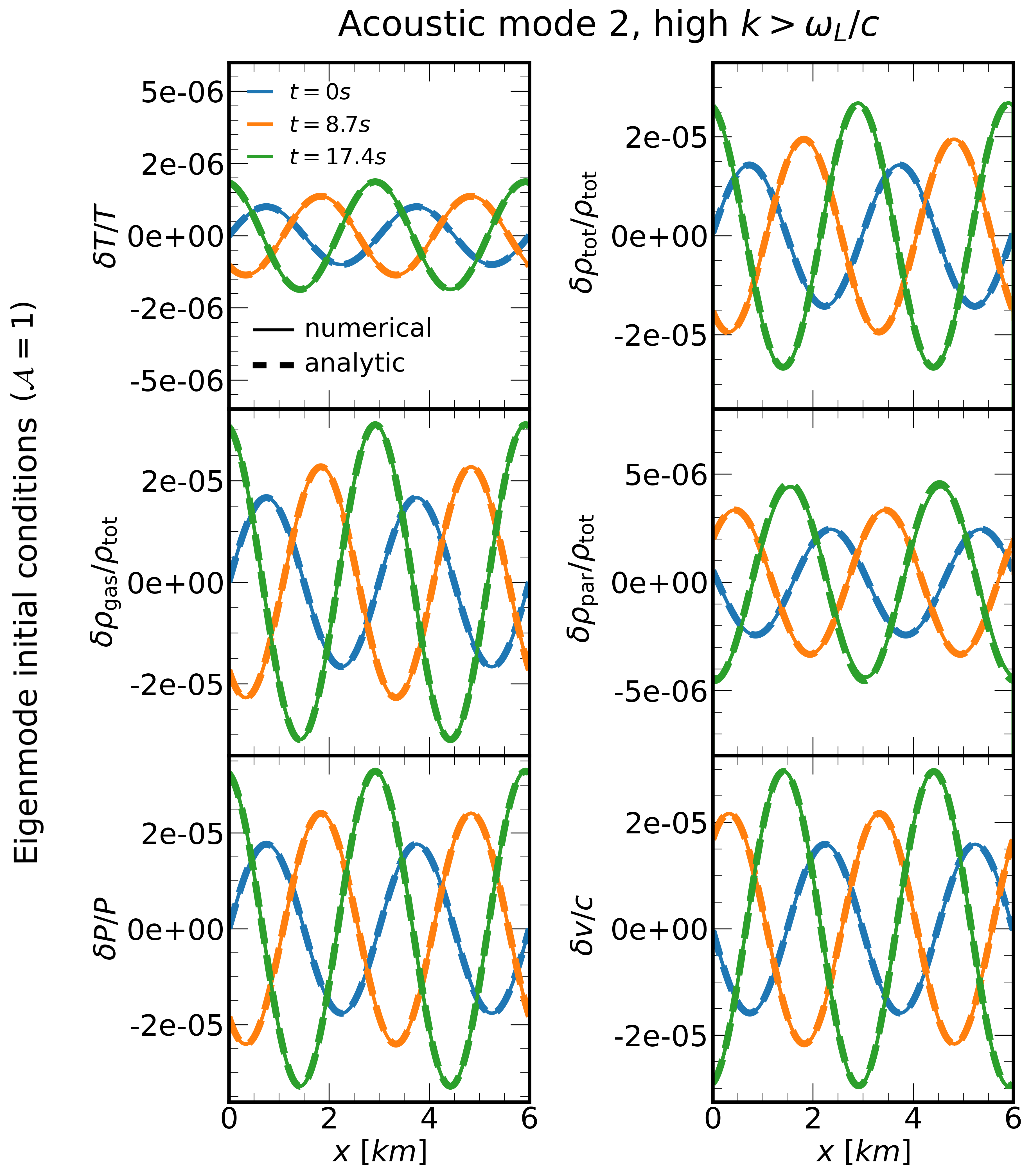}
    \caption{\small 
    Time evolution of mode 2 (sound wave propagating in the positive $x$ direction) at high $k$, with numerical (thin solid, \S\ref{subsec:ot_eigen}) and analytic (thick dashed, \S\ref{subsec:acou_eigen}) solutions overplotted.
    Mode 2 grows more slowly than mode 1; compare with Fig.~\ref{fig:highk_mode1}, and note different timestamps. The numerical solution adopts a background particle-to-gas ratio $\rho_{\rm par}/\rho_{\rm gas} = 10^{-5}$, and the analytic curves are computed in the limit $\rho_{\rm par}/\rho_{\rm gas} \ll 1$.
    }
    \label{fig:highk_mode2}
\end{figure}

\subsubsection{Acoustic eigenmode behavior} \label{subsec:acou_eigen}
At high $k$, $\omega_{\rm Re} = \pm \sqrt{\Gamma} ck$ and $\omega_{\rm Im} = (\Gamma-1-1/a)\omega_L/(2\Gamma) \ll ck$. We follow the same procedure as above to solve (\ref{eq:matrix1}) for the eigenvector:
\begin{align} \label{eq:eigen2}
ck \gg \omega_L: \,\,\,\, \mathbf{E}_{2,3} = \begin{bmatrix} \delta \rho_{\rm tot}/\rho \\ \delta \rho_{\rm gas}/\rho \\ \delta v /c \\ \delta T / T \\ \delta P / P
\end{bmatrix} = 
\begin{bmatrix}
[(1+a)/\Gamma] \left[ 1 \mp i(\Gamma-1-1/a)\Gamma^{-3/2} \omega_L/(ck) \right] \\ a \\  \pm [(1+a)/\sqrt{\Gamma}] \left[ 1 \mp (i/2)(\Gamma-1-1/a)\Gamma^{-3/2} \omega_L/(ck)\right] \\ 1 \\ 1+a
\end{bmatrix}
\delta T/T \,.
\end{align}
The mode behaves like an adiabatic sound wave whose growth can be understood via the work integral (e.g. \citealt{1982ApJ...261..301B}):
\begin{align} \label{eq:work}
+\oint P dV \propto -\oint \frac{P \,d\rho_{\rm tot}}{\rho_{\rm tot}^{2}} \propto -\oint \delta P \frac{\partial \delta \rho_{\rm tot}}{\partial t} dt 
\end{align}
which measures, over one wave period, the work done by a fluid parcel of unit mass to increase the mode kinetic energy, where the parcel volume $V \propto 1/\rho_{\rm tot}$, $P = P_0 + \delta P$, and $\rho_{\rm tot} = \rho_{\rm tot,0} + \delta \rho_{\rm tot}$.\footnote{The work integral over one wave cycle cannot be evaluated for the stationary mode which has no period. Still, it is evident from (\ref{eq:eigen1}) that a given fluid parcel in the stationary mode continuously does positive work on its surroundings: $+P dV \propto -\delta P \, \partial (\delta \rho_{\rm tot})/\partial t > 0$, whether the parcel is in a pressure peak or trough.} If $\oint P dV > 0$, the mode increases in kinetic energy, i.e., the wave amplifies. Suppose $\delta P \propto \cos (kx -\omega_{\rm Re} t)$; then from (\ref{eq:eigen2}), $\delta \rho_{\rm tot} \propto \cos (kx - \omega_{\rm Re} t \mp \varepsilon)$ where $\varepsilon \sim \omega_L/(ck) \ll 1$ is a phase lag. Then $\partial \delta \rho_{\rm tot}/\partial t \propto \omega_{\rm Re} \sin (kx - \omega_{\rm Re}t \mp \varepsilon) \propto \pm k \sin (kx - \omega_{\rm Re}t \mp \varepsilon)$,  and from (\ref{eq:work}) the work integral $\oint P dV \propto k \sin \varepsilon \propto \omega_L > 0$. Thus independent of $k$ in this high-$k$ limit, and regardless of the wave direction, the wave gains energy.

In more physical detail, for an oscillatory mode to gain energy every cycle period, there must be a phase lag $\varepsilon$ such that when a fluid parcel attains an extremum in pressure (either $\max \delta P > 0$ or $\min \delta P < 0$), its volume is still changing (either expanding $\partial \delta \rho_{\rm tot}/\partial t < 0$ for $\delta P > 0$, or contracting $\partial \delta \rho_{\rm tot}/\partial t > 0$ for $\delta P < 0$). In this way positive contributions are  made to the work integral (\ref{eq:work}). How does this phase lag between pressure and volume arise for our system? According to the energy eq.~(\ref{eq:energy}), 
\begin{align} \label{eq:e_interp}
 \left( P + \rho_{\rm gas} L_{\rm vap} \right) \nabla \cdot \mathbf{v}  = \mathcal{H} -4\sigma T^4 \kappa_{\rm par} \rho_{\rm par} \equiv \mathcal{Q} \,\,\,\,\,\,{\rm at \, max \, or \, min\,\,}\delta P \,,
\end{align}
since $DT/Dt = 0$ when pressure reaches an extremum (pressure and temperature are always in phase on the co-existence curve; see any of the eigenvectors). 
We see that at $\max \delta P$, the energy to drive volume expansion ($\nabla \cdot \mathbf{v} > 0$)  can come from decreasing the particle density $\rho_{\rm par}$, so that the background heating term $\mathcal{H}$ exceeds radiative cooling ($\mathcal{Q} > 0$). Indeed, for the high-$k$ acoustic mode, the perturbed particle density
\begin{align}
 \frac{\delta \rho_{\rm par}}{\rho} = \frac{\delta \rho_{\rm tot} - \delta \rho_{\rm gas}}{\rho} = \left[ \frac{(1-\Gamma) a +1}{\Gamma} \mp i \frac{ (\Gamma-1-1/a)(1+a)}{\Gamma^{5/2}} \frac{\omega_L}{ck} \right] \frac{\delta T}{T}  \,\,\,\,\,\, ( ck \gg \omega_L ) \label{eq:drhopar2}
\end{align}
is $< 0$ at $\max \delta T$ (in the square brackets above, the real part dominates and is $<0$ for our fiducial $\Gamma \simeq 1.2$ and $a \simeq 17$). To leading order, particles evaporate as the fluid gets hotter under compression, and the consequent reduction in radiative cooling allows $\mathcal{H}$ to energize the mode at the moment of maximum pressure.

The same energy boost occurs at $\min \delta P < 0$, but for the opposite reason; now $\mathcal{Q} < 0$ because there are more particles, causing radiative losses to dominate and the parcel to shrink ($\nabla \cdot \mathbf{v} < 0$). This cooling phase of the work cycle can occur even if $\mathcal{H}=0$. In principle, from (\ref{eq:e_interp}), all that is needed for a parcel to contract relative to its surroundings is for it to have more particles, which radiate more. Thus we argue that acoustic modes in a medium without background heating can still destabilize, not because they pick up more energy during the high-pressure phase of their cycle, but because they lose more energy during the low-pressure phase.


In the low-$k$ limit, waves are quickly cooled by radiation and behave nearly isothermally:
\begin{align} \label{eq:eigen3}
ck \ll \omega_L:\,\,\,\, \mathbf{E}_{2,3} = \begin{bmatrix} \delta \rho_{\rm tot}/\rho \\ \delta \rho_{\rm gas}/\rho \\ \delta v /c \\ \delta T / T \\ \delta P / P
\end{bmatrix} = 
\begin{bmatrix}
a \left[ 1 \mp i(\Gamma-1-1/a)(1+1/a)^{-1/2} ck/\omega_L \right] \\ a \\  \pm \sqrt{a(1+a)} \left[ 1 \mp (i/2)(\Gamma-1-1/a)(1+1/a)^{-1/2}ck/\omega_L \right] \\ 1 \\ 1+a
\end{bmatrix}
\delta T/T \,.
\end{align}
The work integral now scales as $-\oint \delta P (\partial \delta \rho_{\rm tot}/\partial t) dt \propto k \sin [(\Gamma-1-1/a)(1+1/a)^{-1/2}ck/\omega_L] \propto k^2 > 0$, recovering the growth rate $\omega_{\rm Im} \propto k^2$. A rough description for how low-$k$ waves amplify is that there are phase lags of order $\varepsilon \sim ck/\omega_L$ 
 between variables that increase the mode amplitude by a fractional amount $\varepsilon$ every wave period $1/(ck)$; then the wave $e$-folding time is $[1/(ck)]/\varepsilon \sim \omega_L/(c^2k^2)$.
 
According to the low-$k$ eigenmode (\ref{eq:eigen3}), $\delta \rho_{\rm par} = \delta \rho_{\rm tot} - \delta \rho_{\rm gas} = 0$ when $\delta T$ is maximized ($\delta \rho_{\rm par}$ and $\delta T$ are exactly $\pi/2$ out of phase according to eq.~\ref{eq:eigen3}). This is an asymptotic result. In reality, 
to explain how low-$k$ waves grow, we should have $\delta \rho_{\rm par} < 0$ ($>0$) when $\delta T$ is maximized (minimized). There must be a small difference between the real components of $\delta \rho_{\rm tot}$ and $\delta \rho_{\rm gas}$ that our asymptotic expressions in (\ref{eq:eigen3}) do not capture.

\section{Numerical experiments in linear stability}\label{sec:numexp}

Here we solve the perturbation equations numerically, staying in the linear regime and in 1D, but allowing disturbances to deviate from sinusoids. The equations we solve are more primitive and general forms of the linearized perturbation equations in \S\ref{sec:scientific}:
\begin{align}
    \label{eq:ot_rhotot}
    \frac{\partial\delta\rho_{\rm tot}}{\partial t} =& -\rho_{\rm tot} \frac{\partial \delta v}{\partial x}\\ 
    \label{eq:ot_rhopar}
    \frac{\partial\delta\rho_{\rm par}}{\partial t} =&-\rho_{\rm tot} \frac{\partial \delta v}{\partial x}- \frac{{d\rho_{{\rm sat}}}}{dT}\frac{\partial \delta T}{\partial t}-\frac{d^2\rho_{\rm sat}}{dT^2}\frac{dT}{dt}\delta T \\
    \delta \rho_{\rm gas} =&\ \delta \rho_{\rm tot}-\delta \rho_{\rm par}\\
    \label{eq:ot_mtm}
    \frac{\partial {\delta v}}{\partial t}=& -\frac{1}{\rho_{\rm tot}}\frac{\partial \delta P}{\partial x}\\
    \label{eq:ot_energy}
    \rho_{\rm tot} C \frac{\partial \delta T}{\partial t} =& - C \frac{dT}{dt} \delta \rho_{\rm tot} - P \frac{\partial \delta v}{\partial x} + L_{\rm vap} \left( \frac{\partial \delta \rho_{\rm par}}{\partial t}+\rho_{\rm par} \frac{\partial \delta v}{\partial x} \right) \nonumber\\
    &-4\sigma T^4 \kappa_{\rm par} \,\delta \rho_{\rm par}- 16 \sigma T^3  \rho_{\rm par} \kappa_{\rm par} \,\delta T\\
    \label{eq:ot_eos}
    \delta P =& \frac{k_{\rm B}}{\mu m_{\rm H}}(T \delta\rho_{\rm gas}+\rho_{\rm gas} \delta T) \,.
\end{align}
The background velocity $v=0$, and the background total density $\rho_{\rm tot} =$ constant. Other background quantities $\rho_{\rm par} (t)$, $\rho_{\rm gas} (t)$, $P (t)$, and $T (t)$ are constant in space, and known (explicit) functions of time. While the constant background heating term $\mathcal{H}$ that we introduced in \S\ref{sec:scientific} does not appear explicitly in the above perturbation equations, the effects of $\mathcal{H}$ are implicit in the time evolution (or lack thereof) of background quantities. There are three cases of interest: (i) A fixed background where radiative cooling from background particles ($\rho_{\rm par} > 0$) is balanced by background heating ($\mathcal{H} > 0$; \S\ref{sec:scientific}); (ii) A fixed background
where there is neither background heating ($\mathcal{H} = 0$) nor background cooling ($\rho_{\rm par} = 0$); (iii) A time-varying background where $\mathcal{H} = 0$ and $\rho_{\rm par} > 0$ --- here $T$ will decrease secularly from unbalanced radiative cooling (hence the $dT/dt$ terms in eqs.~\ref{eq:ot_rhopar} and \ref{eq:ot_energy}). We will consider all three cases in \S\ref{subsec:ot_eigen}, \S\ref{subsec:no_par}, and \S\ref{subsec:secular}, respectively.

As in \S\ref{sec:scientific}, we take as small parameters $|\delta T|/T$, $|\delta P|/P$, and $|\delta \rho_{\rm gas}|/\rho_{\rm gas} \ll 1$,   but do not assume $|\delta \rho_{\rm par}| \ll \rho_{\rm par}$. 
Unlike in \S\ref{sec:scientific} (cf.~eqs.~\ref{eq:de} and \ref{eq:dea}), we retain the term $-16\sigma T^3 \rho_{\rm par} \kappa_{\rm par} \, \delta T$ in the energy equation (\ref{eq:ot_energy}) for greater accuracy.

Equations (\ref{eq:ot_rhotot})--(\ref{eq:ot_eos}) are six linear partial differential equations for the six variables $\delta \rho_{\rm tot}$, $\delta \rho_{\rm par}$, $\delta \rho_{\rm gas}$, $\delta P$, $\delta T$, and $\delta v$, all functions of position $x$ and time $t$. They are solved using a staggered leapfrog method \citep{press1992numerical} that is 2nd order accurate in time and space, on an Eulerian grid that resolves a perturbation length scale to a fractional accuracy of $\sim$$10^{-4}$, using timesteps that are typically $\sim$$10^{-5}$ of the total integration duration. We have checked that our solutions have converged with grid cell size and timestep. Periodic boundary conditions are used throughout.

Bounds on $\delta \rho_{\rm par}$ and $\delta \rho_{\rm gas}$ from mass conservation (eqs.~\ref{eq:bounds0}--\ref{eq:bounds2}) are enforced as follows. If at a given timestep the solver's usual algorithm advances $\delta \rho_{\rm par}$ beyond its ``floor'' (\ref{eq:bounds}), then $\delta \rho_{\rm par}$ is re-set to its floor ($-\rho_{\rm par}$), with concomitant re-settings of $\partial \delta \rho_{\rm par}/\partial t$, $\delta T$, and $\delta P$ (the other variables $\delta \rho_{\rm tot}$ and $\delta v$ do not need re-setting). The same flooring procedure is applied to $\delta \rho_{\rm gas}$ (\ref{eq:boundsa}). To check the ``ceiling'' condition (\ref{eq:bounds1}), we track in every grid cell 
\begin{equation}
\label{eq:drhop_phase_changes}
\delta \rho_{\rm par,\ phase\ change\ only}=
\int^{t} \left[ \frac{\partial \delta \rho_{\rm par}}{\partial t}-\left(\frac{\rho_{\rm par}}{\rho_{\rm tot}}\right) \frac{\partial \delta \rho_{\rm tot}}{\partial t}\right] \  dt
\end{equation}
which is a running tally of particle density changes with the contribution from particle transport ($-(\rho_{\rm par}/\rho_{\rm tot}) \partial \delta \rho_{\rm tot}/\partial t$) subtracted off. If (\ref{eq:drhop_phase_changes}) exceeds the ceiling value of $\rho_{\rm gas}$, then all of the background gas has condensed and the calculation is halted. An analogous check is made for the ceiling condition (\ref{eq:bounds1a}). For the calculations shown in this paper, the ceilings are not hit, while the floors sometimes are.

\subsection{Fixed background: Eigenmode evolution ($\rho_{\rm par} > 0$ and $\mathcal{H} > 0$)}\label{subsec:ot_eigen}

As a first test of our numerical solver, we use it to recover the eigenmode evolution derived in \S\ref{sec:scientific}. Accordingly, all background quantities are assumed constant; in particular $dT/dt=0$. 
We adopt $T = 2300$ K, $\kappa_{\rm par} = 2.5$ cm$^2$/g, and $\rho_{\rm gas} = \rho_{\rm sat} (T)$.
We initialize $\delta T/T$ to be the sinusoid (\ref{eq:norm}) with normalization constant $\mathcal{A}=1$. Other perturbation variables are initialized in relation to $\delta T/T$ according to the simplified matrix eq.~(\ref{eq:matrix1}).  

Figures \ref{fig:highk_mode1}, \ref{fig:lowk_mode1}, and \ref{fig:highk_mode2} show the time evolution of perturbations starting from three sets of initial conditions that illustrate, respectively, a high $k > \omega_L/c$ stationary mode, a low-$k$ stationary mode, and a high-$k$ acoustic mode. In each case, the numerical solution compares well with the analytic solution obtained by multiplying initial values by $\exp (-i \omega t)$. Deviations between analytic and numerical solutions are due to technical differences between the calculations. The largest deviations manifest for the high-$k$ stationary mode (Fig.~\ref{fig:highk_mode1}), where the analytic solution (\ref{eq:eigen1}) assumes the limit $\rho_{\rm par}/\rho_{\rm gas} \ll 1$ (eqs.~\ref{eq:dea} and \ref{eq:parnone}), and the 
 numerical solution adopts $\rho_{\rm par}/\rho_{\rm gas} = 0.1$. The latter choice is made to keep $\rho_{\rm par} + \delta \rho_{\rm par} \geq 0$ in the code (the floor condition~\ref{eq:bounds}), since for the high-$k$ stationary mode, $\delta \rho_{\rm par}$ can be negative and grow to especially large magnitude, requiring a commensurately large $\rho_{\rm par}$.
 We have verified that discrepancies between analytic and numerical solutions are eliminated by choosing $\rho_{\rm par} \ll \rho_{\rm gas}$ in the numerical solution, and $\mathcal{A} \ll 1$ to force $|\delta \rho_{\rm par}|/\rho_{\rm par} \ll 1$ at all times (data not shown). To avoid analogous discrepancies for the high-$k$ acoustic mode (Fig.~\ref{fig:highk_mode2}), we set $\rho_{\rm par}/\rho_{\rm gas} = 10^{-5}$, which we can do in this case without violating $\delta \rho_{\rm par} \geq -\rho_{\rm par}$ (even for $\mathcal{A} = 1$) because the acoustic mode grows relatively slowly.


\subsection{Fixed background: No background particles or heating ($\rho_{\rm par} =0$ and  $\mathcal{H}=0$)} \label{subsec:no_par}
We now experiment with a fixed background that has no particles or heating. Relative to the background $\rho_{\rm par} = 0$, the perturbation particle density $\delta \rho_{\rm par}$ can be positive from condensing background gas, but not negative since there are no background particles to evaporate (\ref{eq:bounds}). This discontinuity in behavior cannot be accommodated by the Fourier analysis of \S\ref{sec:scientific}.


Background quantities aside from $\rho_{\rm par} =0$ are the same as those for  \S\ref{subsec:ot_eigen}. Initial perturbations are as follows: $\delta T / T = 10^{-6} \sin [2\pi x/ (0.5 \, {\rm km})]$ (i.e.~$\mathcal{A}=1$ and high $k > \omega_L/c$), $\delta v = 0$, $\delta \rho_{\rm par}=\max(0,-d\rho_{\rm sat}/dT\cdot \delta T)\geq 0$, and $\delta \rho_{\rm gas}=-\delta \rho_{\rm par}$. Our initial $\delta T/T$ is the same as that of the high-$k$ stationary eigenmode of Fig.~\ref{fig:highk_mode1}, but the other perturbation variables do not follow those of any eigenmode. For these input parameters, $\delta \rho_{\rm par}$ hits its floor of 0 and we need to apply the re-setting procedure for the first 2 timesteps of the integration, out of a total of over 40000 steps spanning 4.7 seconds. Re-setting to the floor is hardly needed because the fluid mostly cools when $\mathcal{H}=0$, $\rho_{\rm par} = 0$ and $\delta \rho_{\rm par} \geq 0$ (see the last two terms of the energy eq.~\ref{eq:ot_energy}); eventually $\delta T < 0$ and $\delta \rho_{\rm par} > 0$ everywhere.

Figure~\ref{fig:drhop>=0} shows the time evolution of perturbed quantities for our non-eigenmode initial conditions, and can be compared against Fig.~\ref{fig:highk_mode1} for the stationary eigenmode. Qualitatively, the behaviors shown are similar: mass is transported from regions of low particle density to regions of high particle density. In both figures, the peaks in $\delta \rho_{\rm par}$ grow exponentially with similar $e$-folding times, but the contrast between peaks and troughs is smaller in Fig.~\ref{fig:drhop>=0}: the $\delta \rho_{\rm par} < 0$ troughs of the eigenmode are flattened and made $>0$ when $\rho_{\rm par} = 0$. Figure~\ref{fig:contrast} traces how the peak-to-trough contrast $\max \delta \rho_{\rm par}/\min \delta \rho_{\rm par}$ decreases with time when $\rho_{\rm par}=0$.

Unlike in the eigenmode, phase relationships between $\delta \rho_{\rm par}$, $\delta v$, and $\delta P$ are not fixed. For example, in Fig.~\ref{fig:drhop>=0}, $\max \delta \rho_{\rm par}$ does not always correspond to $\min \delta P$, and fluid velocities $\delta v$ do not always point from high to low $\delta P$ (contrast with Fig.~\ref{fig:highk_mode1}). 

\begin{figure}
    \centering
    \includegraphics[width=0.95\linewidth]{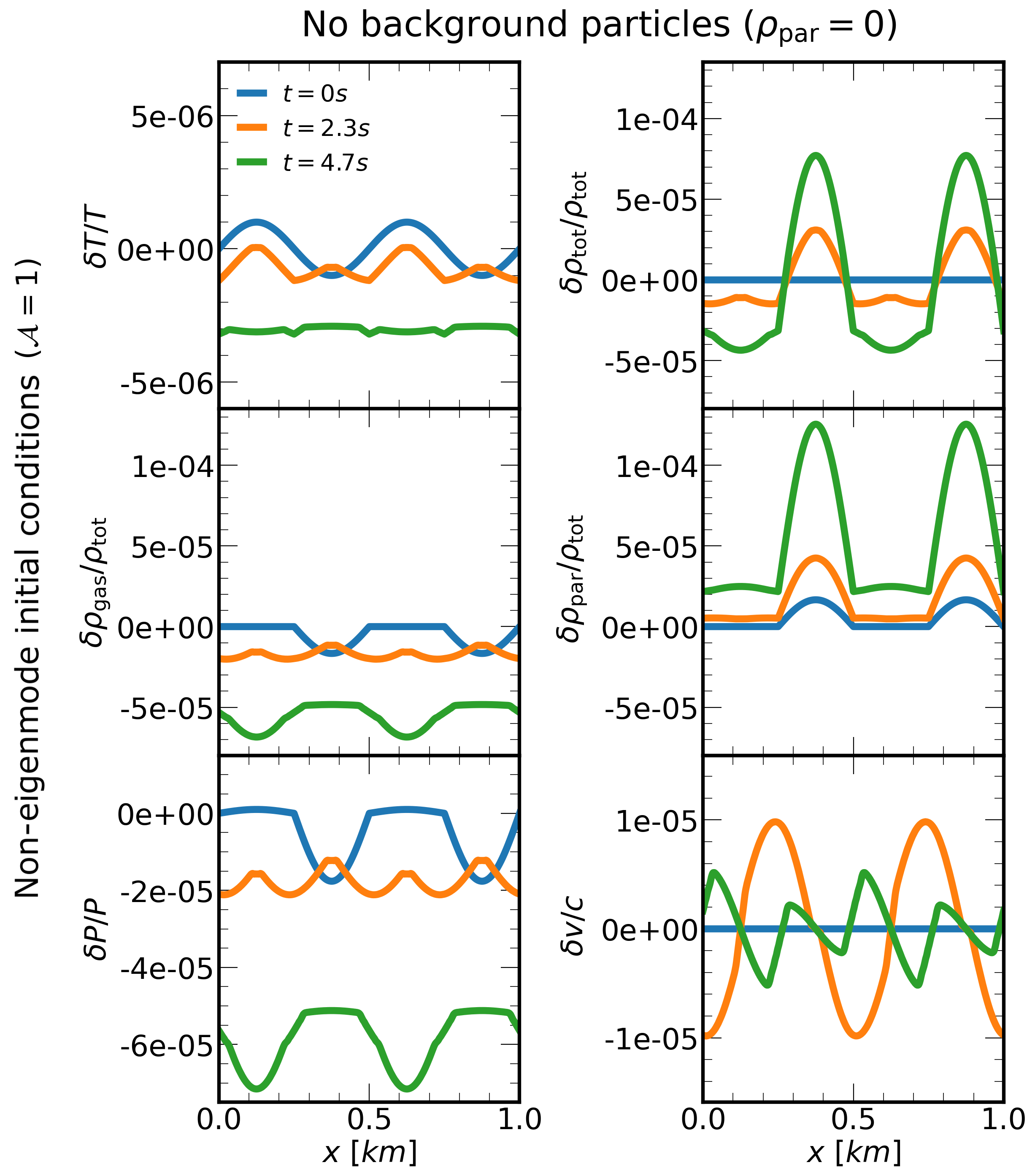}
    \caption{\small Time evolution of perturbations with no background particles ($\rho_{\rm par}=0$) and no background heating ($\mathcal{H}=0$), calculated by numerical solution of eqs.~(\ref{eq:ot_rhotot})--(\ref{eq:ot_eos}) with $\delta \rho_{\rm par} \geq 0$ enforced (\S\ref{subsec:no_par}).
    Input parameters are: $T = 2300$ K, $\rho_{\rm gas} = \rho_{\rm sat}(T)$, $\delta T/T=10^{-6}\sin[2\pi x/ (0.5 \,{\rm km})]$, $\delta v = 0$, $\delta \rho_{\rm par}=\max(0,-d\rho_{\rm sat}/dT\cdot \delta T)$, and $\delta \rho_{\rm gas}=-\delta \rho_{\rm par}$. Since the gas-particle mixture can only cool radiatively with gas condensing into more particles, eventually $\delta T < 0$, $\delta \rho_{\rm gas} < 0$, and $\delta \rho_{\rm par} > 0$ everywhere. As in stationary eigenmode 1, fluid is transported out of initially hot into initially cold regions, amplifying local particle-to-gas ratios above what static condensation would give ($\delta \rho_{\rm par} > -\delta \rho_{\rm gas}$ as opposed to $\delta \rho_{\rm par} = -\delta \rho_{\rm gas}$). Compared to the eigenmode, however, $\max \delta \rho_{\rm par}$ grows more slowly here, and transport is not as coherent; the peaks and troughs of $\delta P$ and $\delta v$ change in position with passing time (contrast with Fig.~\ref{fig:highk_mode1}). Consequently, the difference between particle-rich ($\max \delta \rho_{\rm par}$) and particle-poor ($\min \delta \rho_{\rm par}$) regions diminishes (see also Fig.~\ref{fig:contrast}).
    }
    \label{fig:drhop>=0}
\end{figure}

\begin{figure}
    \centering
    \includegraphics[width=0.6\linewidth]{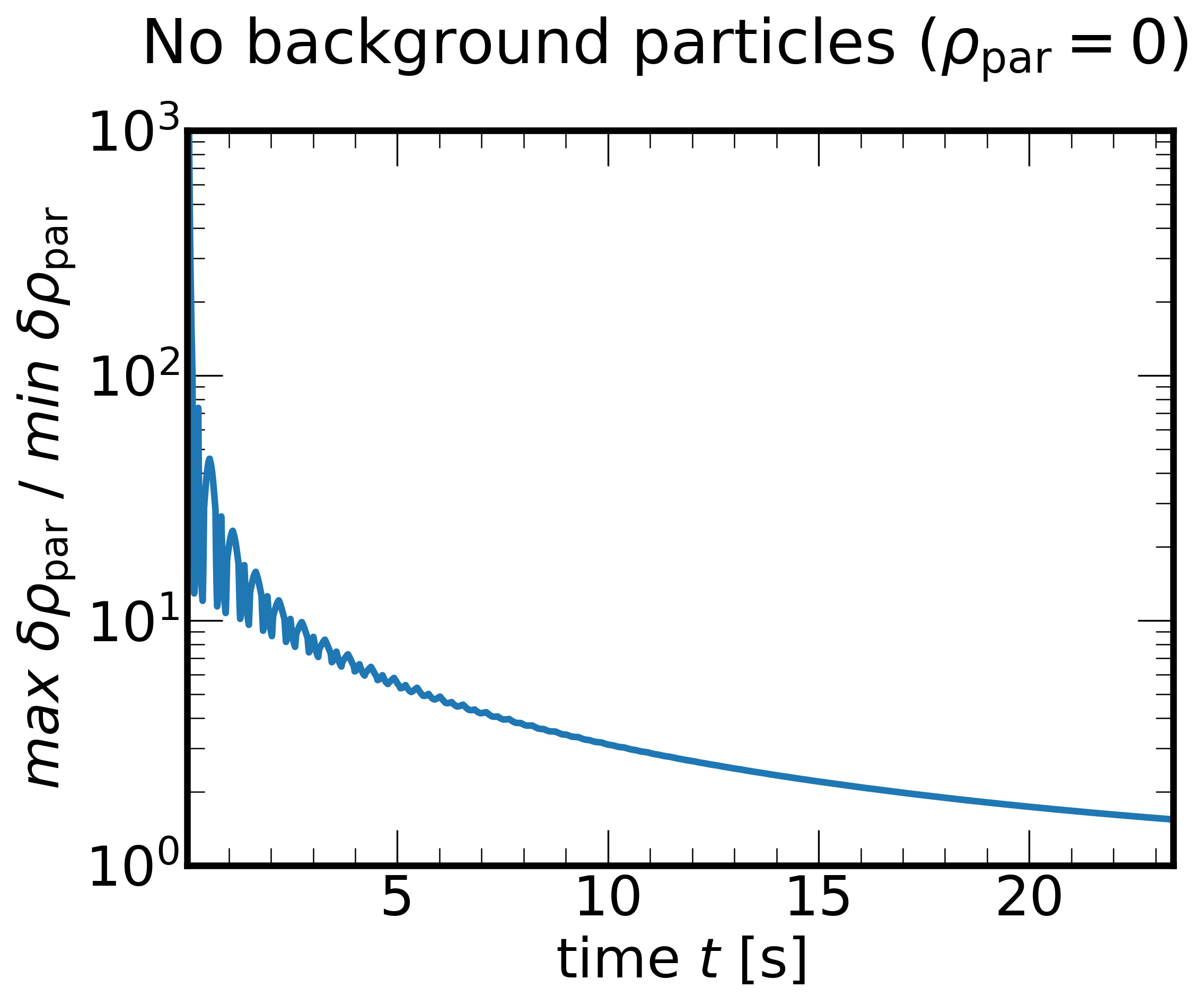}
    \caption{\small Time variation of the contrast between particle-rich and particle-poor regions, in the case where there are no background particles ($\rho_{\rm par}=0$) and no background heating ($\mathcal{H}=0$). Although both $\max \delta \rho_{\rm par}$ and $\min \delta \rho_{\rm par}$ grow exponentially fast (see companion Fig.~\ref{fig:drhop>=0}), the difference between them diminishes, and the medium becomes increasingly uniform.
    }
    \label{fig:contrast}
\end{figure}

\subsection{Time-varying background: A secularly cooling medium ($\rho_{\rm par}>0$ and $\mathcal{H}=0$)} \label{subsec:secular}

\begin{figure}
    \centering
    \includegraphics[width=0.49\linewidth]{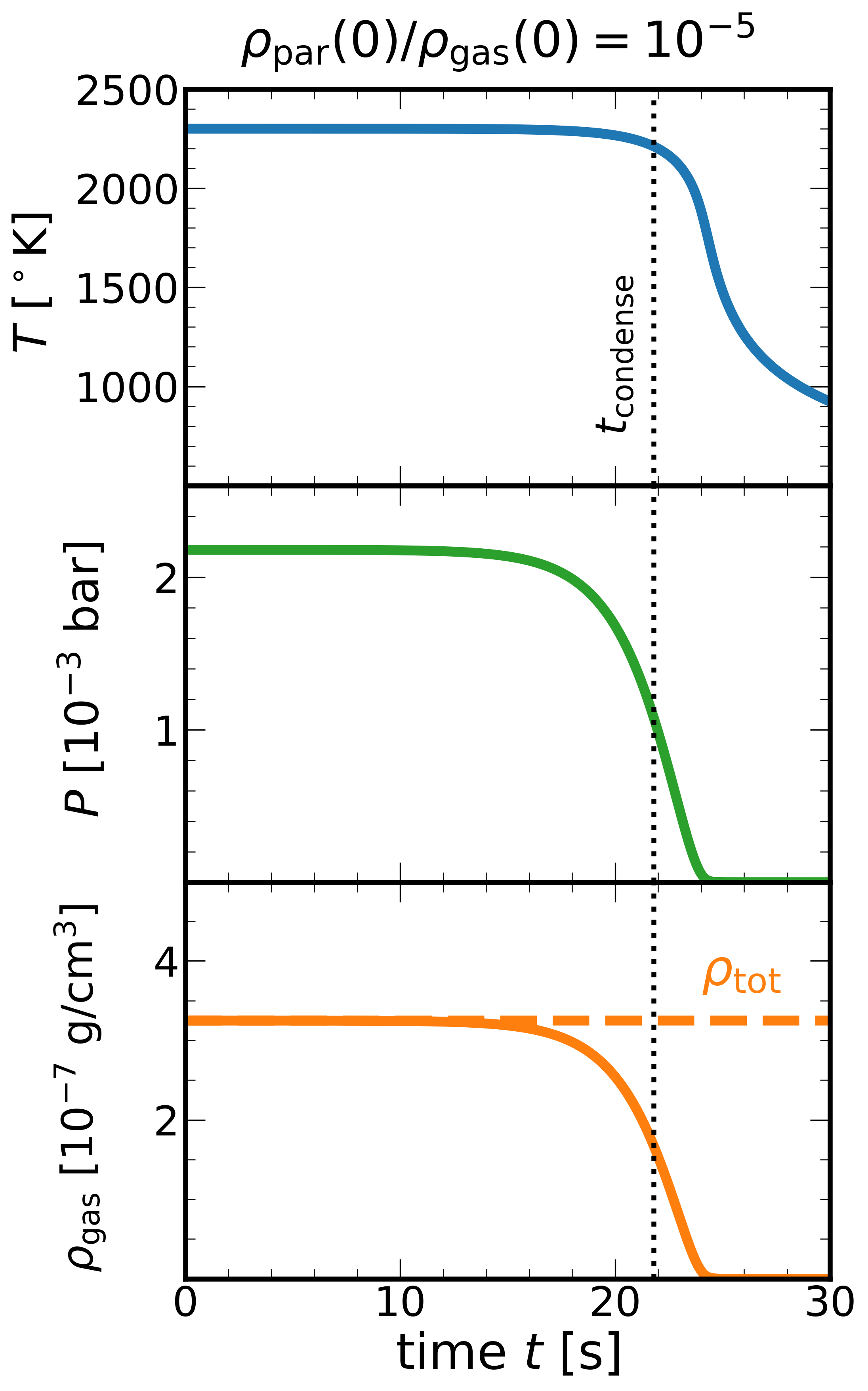}
    \hfill
    \includegraphics[width=0.49\linewidth]{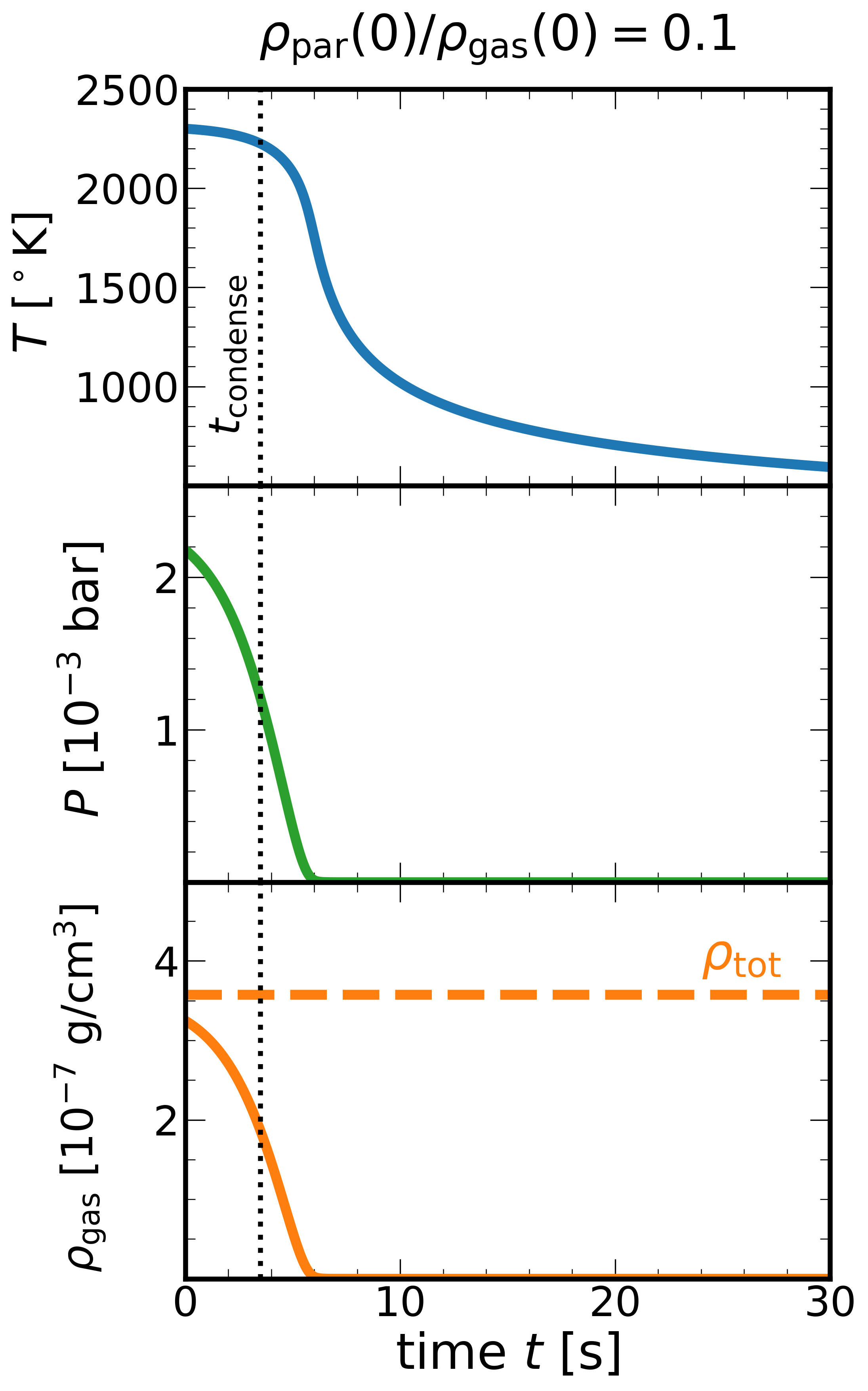}
    \caption{\small When there is a non-zero initial background particle density ($\rho_{\rm par}(t=0)>0$), but no heating ($\mathcal{H}=0$), background quantities $T$, $P$, and $\rho_{\rm gas}$ decrease secularly as the gas-particle mixture radiatively cools, and gas condenses into particles. Dotted vertical lines mark $t_{\rm condense}$ when roughly half of the gas has condensed, as estimated by (\ref{eq:t_cond}). This condensation time is only logarithmically sensitive to the initial particle-to-gas ratio labeled above each set of plots. 
    In the optically thin limit, the rate of temperature decrease scales with $\rho_{\rm par}T^4$ (eq.~\ref{eq:T_bg}), which reaches its maximum near $t_{\rm condense}$ because of increasing $\rho_{\rm par}$, and later falls because of decreasing $T$. 
    Pressure and gas density under saturated conditions are exponentially sensitive to temperature and drop steeply.}
    \label{fig:secular_bg}
\end{figure}
We finally experiment with a background state having a seed particle density ($\rho_{\rm par}>0$) but no heating ($\mathcal{H}=0$). Such a background, assumed spatially uniform and motionless, cools secularly according to eq.~(\ref{eq:energy}) with all $\nabla\cdot$ terms zeroed out:
\begin{equation}
    \label{eq:T_bg}
    \frac{dT}{dt}=-\frac{4\sigma T^4 \rho_{\rm par}\kappa_{\rm par}}{\rho_{\rm tot}C+L_{\rm vap}(d\rho_{\rm sat}/dT)} \,.
\end{equation}
For a given initial temperature $T(0) = 2300$ K, initial gas density $\rho_{\rm gas}(0)=\rho_{\rm sat}(T(0))$, and initial particle-to-gas ratio $\rho_{\rm par}(0)/\rho_{\rm gas}(0)$, eq.~(\ref{eq:T_bg}) is solved as an ordinary differential equation for $T(t)$ with $\rho_{\rm par}=\rho_{\rm tot}-\rho_{\rm sat}(T)$ and $\rho_{\rm tot} = \rho_{\rm gas} + \rho_{\rm par} =$ constant. Remaining parameters are set to fiducial values ($C = 8 \times 10^6$ erg/g/K, $L_{\rm vap} = 3 \times 10^{10}$ erg/g, $\kappa_{\rm par}=2.5$ cm$^2$/g). From $T(t)$ we obtain $P(t)=P_{\rm sat}(T)$ and $\rho_{\rm gas}(t)=\rho_{\rm sat}(T)$.

Figure~\ref{fig:secular_bg} shows the time evolution of background quantities for two initial particle-to-gas ratios $\rho_{\rm par}(0)/\rho_{\rm gas}(0) = \{10^{-5}, 10^{-1} \}$. After a nearly isothermal phase when cooling is slow because particle densities are low, 
the temperature, pressure, and gas density drop precipitously around a time 
\begin{equation}
    \label{eq:t_cond}
   t_{\rm condense}\sim \frac{L_{\rm vap}}{4\sigma T(0)^4\kappa_{\rm par}}\ln \left(\frac{\rho_{\rm tot}/2}{\rho_{\rm par}(0)}\right) 
\end{equation}
when the bulk of the background medium condenses. Equation (\ref{eq:t_cond}) is an order-of-magnitude estimate derived as follows. For a given mass in particles to $e$-fold, radiation must carry away the latent heat of a gas mass comparable to the particle mass. The time for radiation to do so, $L_{\rm vap}/(4\sigma T(0)^4 \kappa_{\rm par})$, is independent of mass or length scale under our assumption that radiation escapes freely. The logarithm in $t_{\rm condense}$ counts the number of particle mass $e$-foldings needed to condense half the medium, and explains why in Fig.~\ref{fig:secular_bg} the timescale for background evolution depends only weakly on $\rho_{\rm par}(0)/\rho_{\rm gas}(0)$.

We substitute $T(t)$, $P(t)$, $\rho_{\rm gas}(t)$, and $\rho_{\rm par}(t) = \rho_{\rm tot}-\rho_{\rm gas}(t)$ into eqs.~(\ref{eq:ot_rhotot})-(\ref{eq:ot_eos}) and use our leapfrog integrator to solve for the evolution of small perturbations atop the time-varying background. 
Figure~\ref{fig:secular_eig} shows the evolution of perturbations using the same high-$k$ stationary eigenmode initial conditions as in Figure~\ref{fig:highk_mode1}. Qualitatively, the evolutions are similar. Background quantities change by only order-unity factors over the time range plotted, and the initial perturbation is small enough that even as the mode grows, neither the floor nor ceiling on $\delta \rho_{\rm par}$ is reached. At the same time $t = 4.7$ s, $\delta \rho_{\rm par}/\rho_{\rm tot}$ is higher in Figure~\ref{fig:secular_eig} than in Figure~\ref{fig:highk_mode1} by a factor of 2. We attribute the faster particle growth rate to the reduction in latent heating as the background gas density declines. One way to see this is to re-derive the stationary mode growth rate in the high-$k$ limit, now taking care to distinguish  between $\rho_{\rm gas}$ and $\rho_{\rm tot}$:
\begin{equation} \label{eq:omg_insta}
\omega_1 = \frac{+ia\omega_T}{b+ \ell  \rho_{\rm gas}/\rho_{\rm tot}} \,.
\end{equation}
The factor of $\rho_{\rm gas}/\rho_{\rm tot}$ is set to unity in the
less general eq.~(\ref{eq:stat1}). Here in (\ref{eq:omg_insta}), $\omega_1$ as $\rho_{\rm gas}$ decreases. Note how $\rho_{\rm gas}$ multiplies against the latent heat parameter $\ell$ (see also the term $\propto L_{\rm vap} \rho_{\rm gas}$ in the master energy eq.~\ref{eq:energy}). 
\begin{figure}
    \centering
    \includegraphics[width=0.95
    \linewidth]{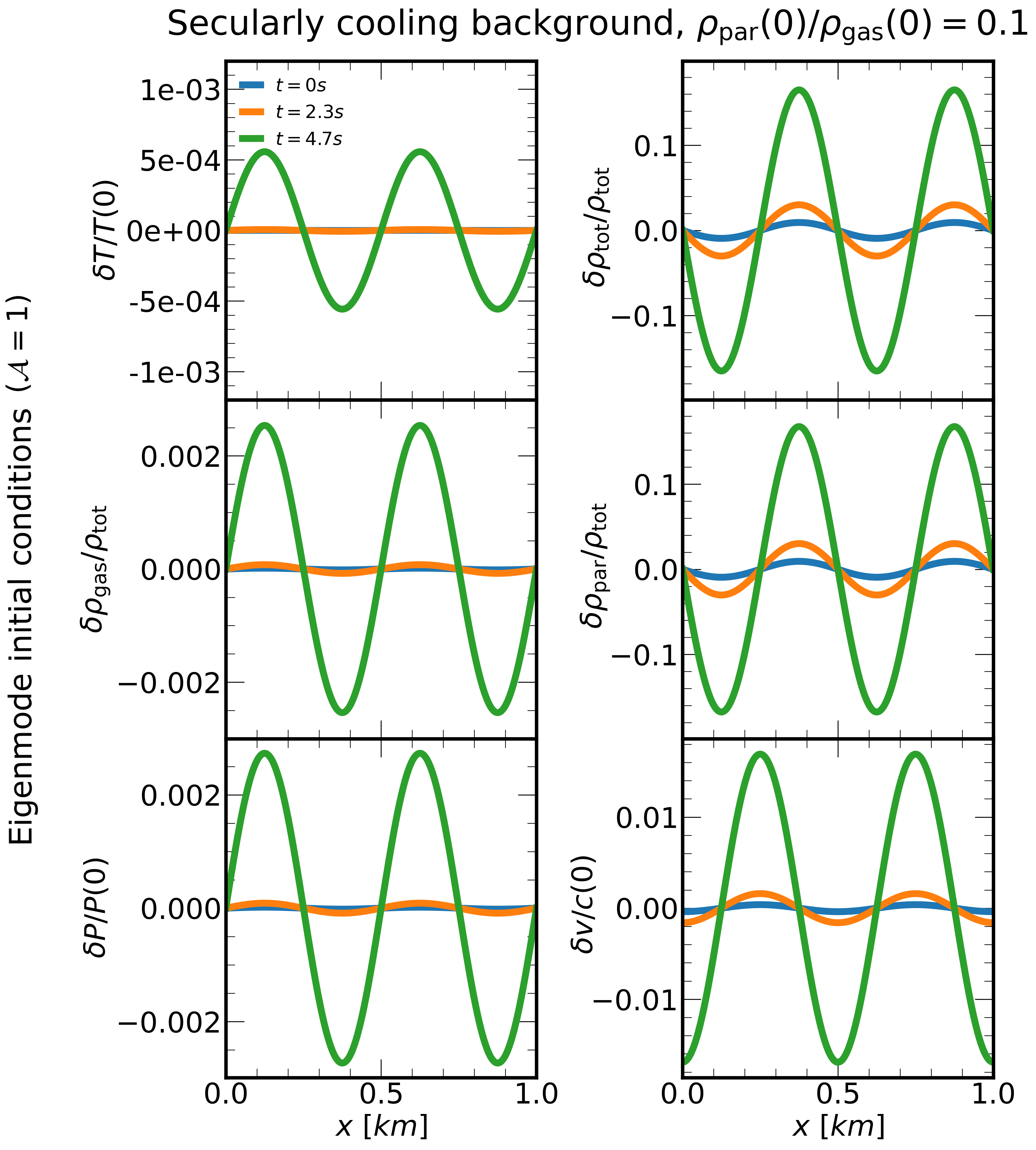}
    \caption{\small Numerical evolution of perturbations on top of a secularly cooling background ($\mathcal{H}=0$, and initial particle-to-gas ratio $\rho_{\rm par}(0)/\rho_{\rm gas}(0) = 0.1$). See Fig.~\ref{fig:secular_bg}, right panel, for how background quantities evolve. Initial perturbations at $t=0$ are identical to those of the high-$k$ stationary eigenmode in Fig.~\ref{fig:highk_mode1}. Comparing the results here for a time-varying background with those in Fig.~\ref{fig:highk_mode1} for a fixed background, we see that perturbations grow similarly --- qualitatively the evolution is that of the high-$k$ stationary eigenmode, with $|\delta \rho_{\rm par}| \gg |\delta \rho_{\rm gas}|$. Perturbations grow faster here as the background gas density $\rho_{\rm gas}$ decreases (eq.~\ref{eq:omg_insta}).}
    \label{fig:secular_eig}
\end{figure}

As a second experiment, we start with non-eigenmode initial conditions: $\delta T/T=7\times 10^{-4}\sin[2\pi x/ (0.5 \,{\rm km})]$ (i.e. $\mathcal{A}=700$ and high $k>\omega_L/c$), $\delta v = 0$, $\delta \rho_{\rm par}=\max(-\rho_{\rm par}(0),d\rho_{\rm sat}/dT\cdot \delta T)$, and $\delta \rho_{\rm gas}=-\delta \rho_{\rm par}$. Background initial conditions are the same as those above except $\rho_{\rm par}(0)/\rho_{\rm gas}(0)=10^{-2}$. These inputs are similar to those of \S\ref{subsec:no_par} except for the larger temperature perturbation and non-zero initial background particle density.

Figures \ref{fig:secular_noneig} and \ref{fig:secular_noneig_ptg} show the evolution from this non-eigenmode experiment. The initially large perturbation amplitude ($\mathcal{A}=700$) and relatively small $\rho_{\rm par}(0)$ cause $\delta \rho_{\rm par}$ to hit the floor of $-\rho_{\rm par}$ at $t=0$: the troughs in $\delta \rho_{\rm par}$ start flattened. Afterward, the $-\rho_{\rm par}$ floor (tracked by horizontal dashed lines in Fig.~\ref{fig:secular_noneig}) becomes more negative from condensing background gas, freeing the $\delta \rho_{\rm par}$ troughs to become more negative as well (contrast with Fig.~\ref{fig:drhop>=0}). Mass is transported at relatively large velocities out of particle troughs and into particle crests, and eventually at $t = 8.5$ s, the particle trough again hits the floor. The creation of a particle void  ($\delta \rho_{\rm par} + \rho_{\rm par} = 0$) at this time can be seen more directly in the companion Fig.~\ref{fig:secular_noneig_ptg}, which tracks the evolution of the total particle-to-gas ratio at the location of minimum $\delta \rho_{\rm par}$.

\begin{figure}
    \centering
    \includegraphics[width=0.95\linewidth]{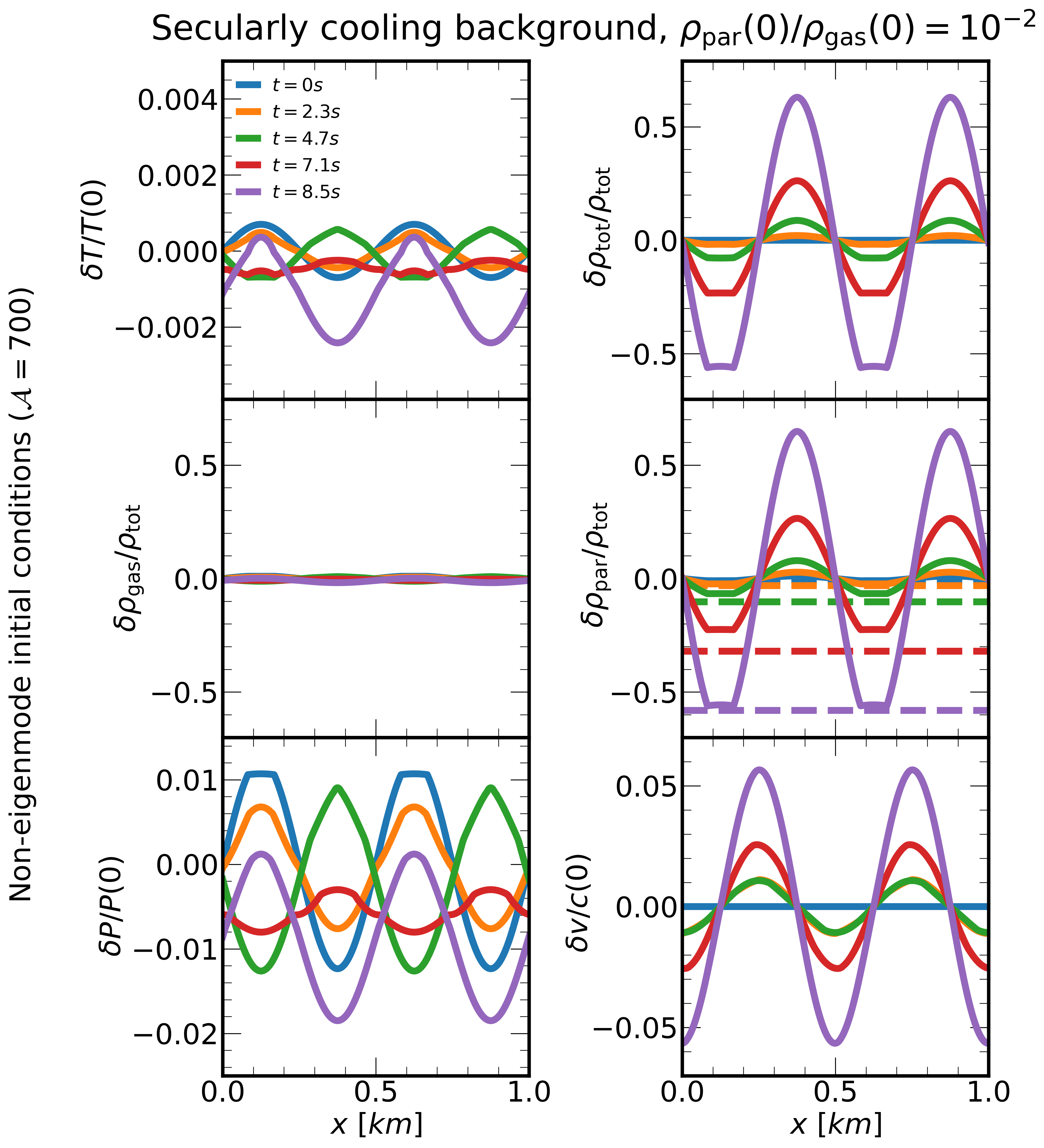}
    \caption{\small Another numerical experiment in perturbing a background that secularly cools ($\mathcal{H}=0$), now with an initial background particle-to-gas ratio of $\rho_{\rm par}(0)/\rho_{\rm gas}(0) = 10^{-2}$, and non-eigenmode initial conditions:
    $\delta T/T=7 \times 10^{-4}\sin[2\pi x/ (0.5 \,{\rm km})]$, $\delta v = 0$, $\delta \rho_{\rm par}=\max(-\rho_{\rm par}(0),d\rho_{\rm sat}/dT\cdot \delta T)$, and $\delta \rho_{\rm gas}=-\delta \rho_{\rm par}$.
Dashed horizontal lines in the middle right panel mark the $-\rho_{\rm par}(t)/\rho_{\rm tot}$ ``floor'', i.e.~the lower bound on $\delta \rho_{\rm par}/\rho_{\rm tot}$ from mass conservation; the condition $\rho_{\rm par} + \delta \rho_{\rm par} \geq 0$ is enforced following a re-setting procedure. Our simulation parameters and initial conditions (in particular a relatively large initial temperature perturbation, $\mathcal{A}=700$) are such that $\rho_{\rm par} + \delta \rho_{\rm par}$ hits zero in initially hot regions, at $t=0$ and $t \simeq 8.5$ s (by which time more than half the background gas has condensed). The particle voids at the end of the simulation are carved out by relatively large velocities $\delta v$. See the companion Fig.~\ref{fig:secular_noneig_ptg}.}
    \label{fig:secular_noneig}
\end{figure}

\begin{figure}
    \centering
    \includegraphics[width=0.8\linewidth]{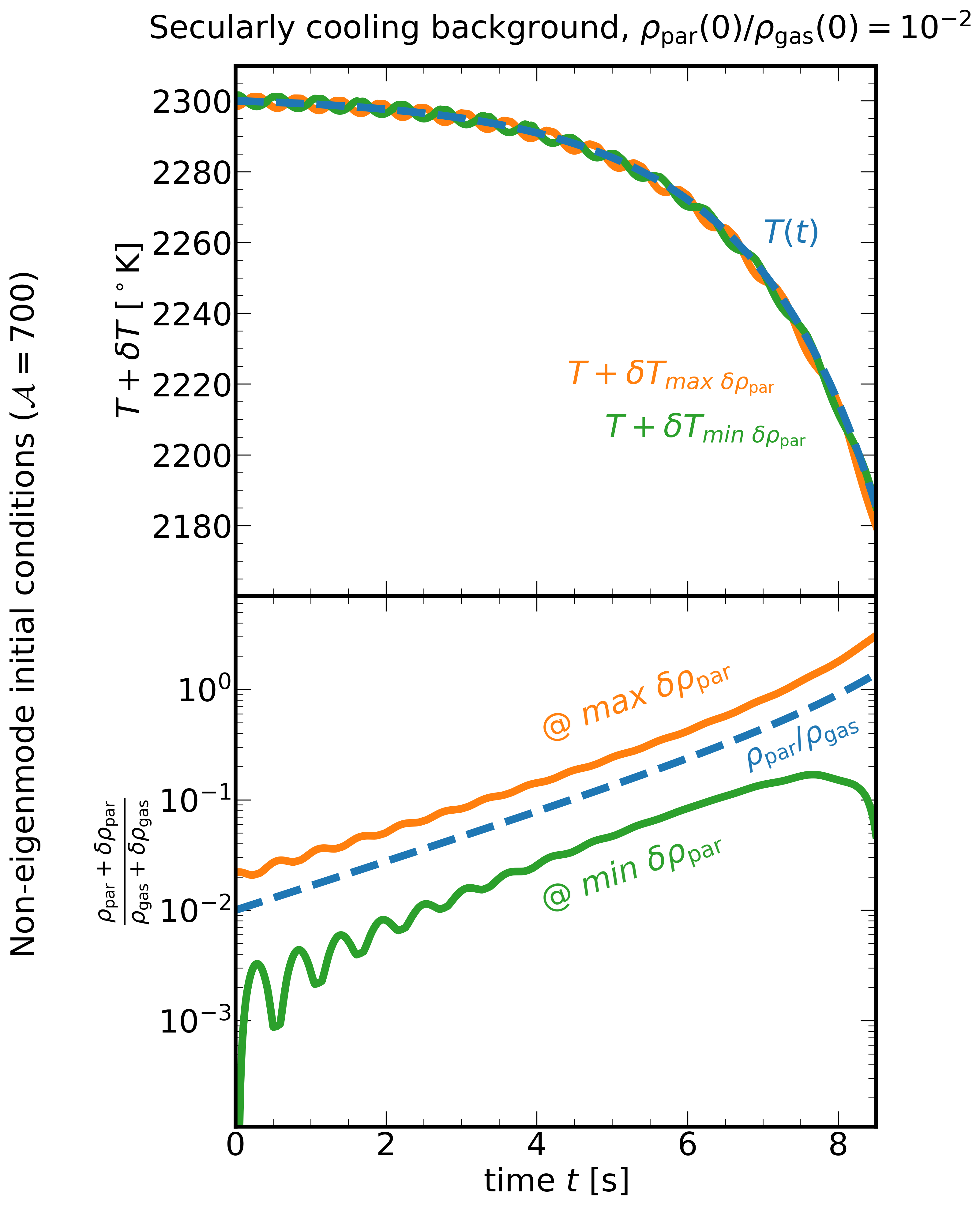}
    \caption{\small Evolution of temperatures and particle-to-gas ratios in the perturbed, secularly cooling fluid simulated in Fig.~\ref{fig:secular_noneig}. {\bf Upper panel:} Temperatures of the background, the location of maximum perturbed particle density (subscript $\max \delta \rho_{\rm par}$), and the location of minimum (most negative) perturbed particle density (subscript $\min \delta \rho_{\rm par}$). The location of $\min \delta \rho_{\rm par}$ varies with time and is not always half a wavelength away from the location of $\max \delta \rho_{\rm par}$ (see Fig.~\ref{fig:secular_noneig}).
    {\bf Lower panel:} Particle-to-gas ratios of the background $\rho_{\rm par}(t)/\rho_{\rm gas}(t)$, the location of maximum perturbed particle density $[(\rho_{\rm par}+\delta \rho_{\rm par})/(\rho_{\rm gas}+\delta \rho_{\rm gas})]_{\max \delta \rho_{\rm par}}$, and the location of minimum perturbed particle density $[(\rho_{\rm par}+\delta \rho_{\rm par})/(\rho_{\rm gas}+\delta \rho_{\rm gas})]_{\min \delta \rho_{\rm par}}$.}
    \label{fig:secular_noneig_ptg}
\end{figure}

\section{Summary and discussion}\label{sec:sum}

We have shown that a hot saturated vapor and its particle condensates
are subject to a linear instability whereby particle overdensities
amplify exponentially. When particles freely radiate their energy to
infinity, the particle-rich get richer and the particle-poor get
poorer --- regions of saturated gas that are overdense in particles
radiate more, thereby cooling and condensing faster. In the fastest
growing mode, clumps and voids grow in place (the mode has zero phase
velocity) with an $e$-folding time  $\omega_L^{-1}$ equal to the time
it takes a perturbation to radiate away its latent heat of
condensation:
$\omega_L^{-1} \simeq L_{\rm vap}/(4\sigma T^4 \kappa_{\rm par} )$ for
latent heat $L_{\rm vap}$ (energy per gas mass), Stefan-Boltzmann constant $\sigma$,
temperature $T$, and opacity $\kappa_{\rm par}$ (cross
section per particle mass). This growth time is independent of
wavelength $2\pi/k$ in the radiation free-streaming limit, and
measured in seconds for mm-sized particles at $T \simeq 2300$
K. Particle densities grow most dramatically if the perturbations are in pressure communication, i.e. if their sound-crossing times $1/(ck)$ are shorter than the growth time 
$\omega_L^{-1}$. In this
high $k > \omega_L/c$ regime, pressure gradients have enough time to
transport mass out of high-temperature, high-pressure, particle-poor
zones into low-temperature, low-pressure, particle-rich zones. For
reference, a cloud of saturated silicate vapor kilometers across
contains the mass equivalent of a solid planetesimal tens of meters in
size.

This radiation-condensation instability is present whether or not the medium is subject to a constant background heating term $\mathcal{H}$. If $\mathcal{H} \neq 0$, then an equilibrium state can be formally defined and perturbed, leading to unstable eigenmodes. The eigenmode analysis is akin to that of Field's (\citeyear{field_1965}) thermal instability; there as here, the dispersion relation for Fourier modes is a cubic equation for wave frequency, with two acoustic modes and a ``thermal condensation'' mode (analogous to our high-$k$, zero phase speed, fast growing mode), all of which can be unstable. If on the other hand $\mathcal{H} = 0$, then no equilibrium can be defined, as the fluid cools secularly from whatever particles are present. Perturbations on top of this time-varying background are still unstable, as we have shown by numerical experiment. Perturbations grow faster as the background gas density decreases and latent heating diminishes. 

Our study assumed a wholly condensable gas, i.e.~a medium composed entirely of silicates and/or metals. Vapor plumes from colliding asteroids are practically H$_2$-free insofar as plume pressures overwhelm nebular pressures (\citealt{choksi_etal_2021}; see also \citealt{campbell_etal_2002}). The solar nebula may have largely dissipated by the time the impacts creating CB/CH chondrites occurred \citep{krot_etal_2005,desch_etal_2023a,desch_etal_2023b}. Wholly condensable, second-generation gas from vaporizing collisions between rocky/icy bodies may also be found in extrasolar debris disks, orbiting young stars and white dwarfs (e.g.~\citealt{marino_etal_2022}; \citealt{swan_etal_2023}; and references therein).

Adding an inert, non-condensable gas like hydrogen to our particle-gas mixture increases its dynamical and thermal inertia, and would be expected to slow, if not stop, growth of perturbations. Unlike the vapor pressure of a condensable gas, the pressure from H$_2$ is not exponentially sensitive to changes in temperature, and would stiffen the fluid and reduce the pressure gradients that otherwise transport mass out of voids and into clumps. \citet{chiang_2024} found that adding H$_2$ to condensing, cavitating bubbles greatly slowed, but did not necessarily halt their collapse. If the radiation-condensation instability can survive the presence of H$_2$, its applicability would widen, from asteroid collisions to potentially snowlines in hydrogen-rich protoplanetary disks. Numerical simulations of disks have shown that meridional flows can transfer water vapor from just inside the water snowline to outside \citep{wang_etal_2025}. As the water vapor cools and re-condenses, it might be subject to clumping by the radiation-condensation instability. A thermal instability on macroscopic disk scales may also cause the CO snowline to vary cyclically \citep{owen_2020}. Though tempered by H$_2$, the radiation-condensation instability might still generate overdensities large enough to trigger other concentration mechanisms, such as the streaming instability and gravitational instability (e.g.~\citealt{li_youdin_2021}).

Our linear instability depends on optically thin radiative cooling: the ability of particle overdensities to shed their energy to infinity. The nonlinear study of cavitating bubbles by \citet{chiang_2024} found a similar requirement: although the bubbles themselves could be optically thick, their surroundings needed to have a lower radiation temperature to serve as an energy sink. To our knowledge, Field's thermal instability only manifests in environments where cooling photons can freely escape, including the solar corona (e.g.~\citealt{brughmans_etal_2022}), the diffuse interstellar medium (e.g.~\citealt{jennings_li_2021}), and the intracluster medium in galaxy clusters (e.g.~\citealt{qiu_etal_2020}).

If instead radiative cooling is treated in the optically thick limit, with the free-streaming term $\propto -T^4 \rho_{\rm par}$ in the energy equation replaced with a diffusive term $\propto \nabla \cdot (\rho_{\rm par}^{-1}\nabla T^4)$ (where $\rho_{\rm par}$ is the particle density), then instability is suppressed, as sound waves are damped by diffusion, including by thermal gas conduction (e.g.~\citealt{field_1965}). 
This presents a problem for applying thermal instability to the vapor plume from colliding asteroids, insofar as the plume may be optically thick \citep{choksi_etal_2021}. The same problem is noted by \citet{chiang_2024}, who suggests that the radiation-condensation instability may be confined to the edges of a debris cloud, or to times when the cloud is more transparent --- either early on when the cloud is too hot for many solids to condense, or later when the cloud has thinned out. Perhaps the plume fragments from the outside in, with clumps and voids forming on the plume's photospheric surface, exposing deeper layers that cool and fragment in turn. Numerical simulations are needed to explore these global, nonlinear scenarios.

Shear flows and turbulence within the impact plume may also interfere with thermal instability.  See 
\citet{balbus_1986} for how thermal instability plays out atop a dynamical flow, and
\citet{robertson_goldreich_2012} for how a turbulent gas evolves upon compression or expansion.

\vspace{0.2in}
\noindent 
We thank Steve Desch, Rixin Li, Francois Tissot, Andrew Youdin, J.J.~Zanazzi, and Shangjia Zhang for discussions. An anonymous referee provided an encouraging and helpful report. We are also grateful to Andrew Ingersoll for reminding us of the application of thermal instability to solar prominences. This work was supported by Berkeley's Esper Larsen,~Jr.~fund, and a Simons Investigator grant.

\software{numpy \citep{numpy_cite},
          scipy \citep{scipy_cite},
          matplotlib \citep{hunter_etal_2007}
          }

\bibliography{rci}
\bibliographystyle{aasjournal}

\end{document}